# Photovoltaic Possibility of $Cu_2SiSe_3$ and $Cu_2SnS_3$ Ternary Chalcogenides: Single Junction to Tandem Architecture

Saptarshi Mandal, Surbhi Ramawat, Sumit Kukreti, and Ambesh Dixit*

Advanced Materials and Device (A-MAD) Laboratory, Department of Physics, Indian Institute of Technology Jodhpur, Jodhpur, Rajasthan, 342030 India

*ambesh@iitj.ac.in

## Abstract

Cu-based ternary chalcogenides are gathering attention for sustainable energy applications due to their reduced complexity compared to quaternary alternatives. We used drift-diffusion modeling to evaluate the feasibility of photovoltaics employing ternary chalcogenide absorbers based on $Cu_2SiSe_3$ and $Cu_2SnS_3$. The device metrics are evaluated by analyzing absorber layer thickness, intrinsic carrier concentration, defect density, and energy band alignment at interfacial junctions. The optimized single-junction $Cu_2SiSe_3$-based device configuration achieves a power conversion efficiency of 18.13%, exhibiting a short-circuit current density of 38 mA/cm² and an open-circuit voltage of 0.64 V. The $Cu_2SnS_3$-based device achieves an efficiency of 15.59%, with a short-circuit current density of 48.8 mA/cm² and an open-circuit voltage of 0.42 V. We examined the impact of the buffer layer on device parameters, uncovering further avenues for performance improvement. Additionally, we simulated a two-terminal tandem solar cell using $Cu_2SiSe_3$ (bandgap: 1.44 eV) in the upper cell to capture photons from the visible spectrum and $Cu_2SnS_3$ (bandgap: 0.91 eV) in the lower cell to absorb from the infrared spectrum. The simulated tandem architecture, featuring a $V_{OC}$ of 1.24 V, a $J_{SC}$ of 24.6 mA/cm², a fill factor (FF) of 79.2%, and an efficiency of 24.1%, markedly surpassed conventional single-junction devices, demonstrating the viability of $Cu_2SiSe_3$-$Cu_2SnS_3$ absorber-based tandem solar cells for next-generation high-efficiency solar technologies.

# 1. Introduction

First-generation silicon-based solar cells currently dominate the photovoltaic market, mostly due to their established manufacturing ecosystem and power conversion efficiency exceeding 26% [1]. Nonetheless, silicon photovoltaics encounter many inherent limitations. The indirect bandgap of silicon results in a comparatively low absorption coefficient, requiring thicker absorber layers to adequately collect sunlight, thereby escalating material consumption and production costs. Furthermore, the manufacturing method is significantly energy-intensive, necessitating high processing temperatures [2,3]. Further performance improvements rely on light management methods, such as surface texturing and anti-reflective coatings, which increase system complexity and expenses. Current research has pivoted toward the development of cost-effective, high-efficiency alternatives to address these restrictions, especially in the field of thin-film solar cells.

Among early thin-film technologies, III-V compound semiconductors have demonstrated exceptional performance, with GaAs-based solar cells reaching 29% efficiency, and GaInP achieving 22% [4,5]. Within the chalcogenide family, CdTe and CIGS technologies have also shown considerable commercial viability, recording efficiencies of 21% [6] and 23.35% [7], respectively. More recent developments in kesterite materials, such as CZTSSe, have achieved efficiencies of 13.45% [8], although these values still fall short of the Shockley-Queisser limit of 33% for single-junction solar cells. Ternary and quaternary chalcogenide compounds have emerged as formidable possibilities owing to their elevated absorption coefficients (~$10^4$ $cm^{-1}$) [9,10], enabling absorber layers that are up to 100 times thinner than those made of silicon. Moreover, these materials often exhibit direct and tunable band gaps, making them ideal for efficient light absorption and energy conversion.

Fabrication difficulties persist, especially in quaternary systems such as CZTS, where the formation of secondary phases poses a major obstacle. During crystal growth, unexpected phases such as ZnS and $Cu_2SnS_3$ (CTS) frequently appear, compromising material purity and device performance [11,12]. Consequently, there is renewed interest in investigating ternary chalcogenides, such as CTS, as more straightforward and controllable alternatives. Despite the past underperformance of CTS-based devices, which attained efficiencies of 4.29% in 2015 [13], 6.7% with Ge-doping in 2016 [14], and 5.2% with Na-doping in 2021 [15], these outcomes signify significant advancement for an emerging material class. Theoretical simulations indicate that optimized CTS devices may exceed 15% efficiency, underscoring their unexploited potential. Moreover, their minimal production expenses and compatibility with flexible substrates make them especially appealing for niche applications, including portable electronics. Furthermore, $Cu_2SiSe_3$, a recently investigated chalcogenide with a bandgap of around 1.42 eV and an intrinsic p-type characteristic, appears to be appropriate for single junction applications due to its earth-abundant components and has also been examined for sustainable energy applications [16,17].

This study investigates the photovoltaic potential of the ternary chalcogenide absorbers $Cu_2SiSe_3$ and $Cu_2SnS_3$ using drift-diffusion modeling. Our principal objective is to evaluate the performance of single-junction absorber layers and predict the potential of a tandem configuration device utilizing $Cu_2SiSe_3$ as the upper cell and $Cu_2SnS_3$ as the lower cell. This work commences with the optimization of absorber thickness, demonstrating that increasing thickness enhances light absorption and device performance to an optimal extent. We subsequently adjust the carrier concentration to identify an optimal range that balances recombination losses and carrier mobility. The influence of bulk and interfacial defects is examined, utilizing defect values that align with authentic material characteristics. Furthermore, we do a comprehensive analysis of the buffer layer, evaluating the effects of its thickness, carrier concentration, and electron affinity. We have also assessed the influence of the back contact's role. We ultimately extend our research to a tandem solar cell configuration using $Cu_2SiSe_3$ as the upper cell and $Cu_2SnS_3$ as the lower cell. Using the complementary optical properties of $Cu_2SiSe_3$, with a bandgap of 1.44 eV for visible light, and $Cu_2SnS_3$, with a bandgap of 0.91 eV for infrared absorption, we demonstrate the feasibility of a high-efficiency, cost-effective tandem configuration suitable for next-generation photovoltaic systems.

## 2. Methodology

In this study, we first propose and model single-junction solar cell architectures employing $Cu_2SiSe_3$ and $Cu_2SnS_3$ as the primary absorber layers, with respective direct band gaps of 1.44 eV and 0.91 eV [18]. We simulated Mo/absorber/CdS/i-ZnO/Al:ZnO device structure. Molybdenum (Mo), featuring a work function of 5.0 eV, is utilized as the back contact electrode. For the buffer layer, cadmium sulfide (CdS), an n-type semiconductor with a wide band gap of 4.24 eV, is selected to facilitate the formation of the critical p-n junction with the p-type absorber. The window layer consists of zinc oxide (ZnO), while aluminium-doped zinc oxide (Al:ZnO) serves as the transparent conducting oxide (TCO), ensuring efficient light transmission and charge collection. **Figure 1** presents a schematic illustration of the device structures for both $Cu_2SiSe_3$ and $Cu_2SnS_3$ and their corresponding energy band alignments, highlighting the direction of carrier flow across the junctions towards their respective electrodes.

The heterojunction between the Cu-based absorber and CdS constitutes the device's fundamental active region, enabling efficient separation and transport of photogenerated carriers. All material and layer parameters used in the simulation are summarized in Table S1 in the supplementary material, and the defect parameters are listed in Table S2 in the supplementary material. The properties of $Cu_2SiSe_3$ are based on predictive data reported in [16], while the parameters for the remaining layers are drawn from established literature or estimated based on reasonable physical assumptions.

Device modeling was conducted using SCAPS-1D version 3.3.11 [19]. It solves the coupled one-dimensional Poisson (equation 1) and continuity (equations 2 and 3) equations for electrons and holes, incorporating interface boundary conditions and the relevant physical mechanisms governing carrier transport. Simulations were performed under standard AM1.5G solar illumination at an intensity of 1000 W/m² and a temperature of 300 K [20].

$$\frac{d}{dx}\left(-\frac{\varepsilon(x)d\Psi}{dx}\right) = q\left[p(x) - n(x) + N_d^+(x) - N_a^-(x) + p_{t(x)} - n_{t(x)}\right] \tag{1}$$

$$\frac{dp_n}{dt} = G_p - \frac{p_n - p_{n0}}{\tau_p} - \frac{p_n \mu_p d\xi}{dx} - \mu_p \xi \frac{dp_n}{dx} + \frac{D_p d^2 p_n}{dx^2} \tag{2}$$

$$\frac{dn_p}{dt} = G_n - \frac{n_p - n_{p0}}{\tau_n} - \frac{n_p \mu_n d\xi}{dx} - \mu_n \xi \frac{dn_p}{dx} + \frac{D_n d^2 n_p}{dx^2} \tag{3}$$

Where $\Psi$ represents the electrostatic potential, $\varepsilon$ represents dielectric permittivity. n and p represent the free electrons and holes, whereas $n_t$ and $p_t$ represent the trapped electrons and holes, q is the electronic charge, and $D$ is the diffusion coefficient. $G$ is the generation rate of charge carriers. $N_d^+$ and $N_a^-$ is the donor and acceptor doping concentration.

For the simulation of the tandem solar cell comprising $Cu_2SiSe_3$ as the top cell and $Cu_2SnS_3$ as the bottom cell absorber. We created the definition files for the top and bottom cells separately and joined them using a script. To ensure an ideal ohmic contact and minimize recombination at the back contact of the top cell, a flat-band condition is assumed. Current matching between the two sub-cells is achieved by optimizing the absorber thickness. The $Cu_2SiSe_3$ absorber thickness is varied from 0.1 µm to 0.5 µm, and the corresponding short-circuit current density ($J_{SC}$) is calculated using the filtered spectrum, given by Equation 4[21 22]:

$$S(\lambda) = S_o(\lambda) \cdot \prod_{m=1}^{M} e^{-\alpha(\lambda)_m d_m} \tag{4}$$

Where λ is the wavelength, $S_0(\lambda)$ represents the AM1.5 Global spectrum, $S(\lambda)$ is the filtered spectrum, $\alpha_m$ is the coefficient of absorption of the m-th layer, and $d_m$ is its thickness. Using the transmitted spectra for various top cell thicknesses, the $J_{SC}$ of the bottom cell is computed for varying $Cu_2SnS_3$ thicknesses. This short-circuit current from both devices is analyzed; a current-matching condition is found, which is then used to simulate the tandem device.

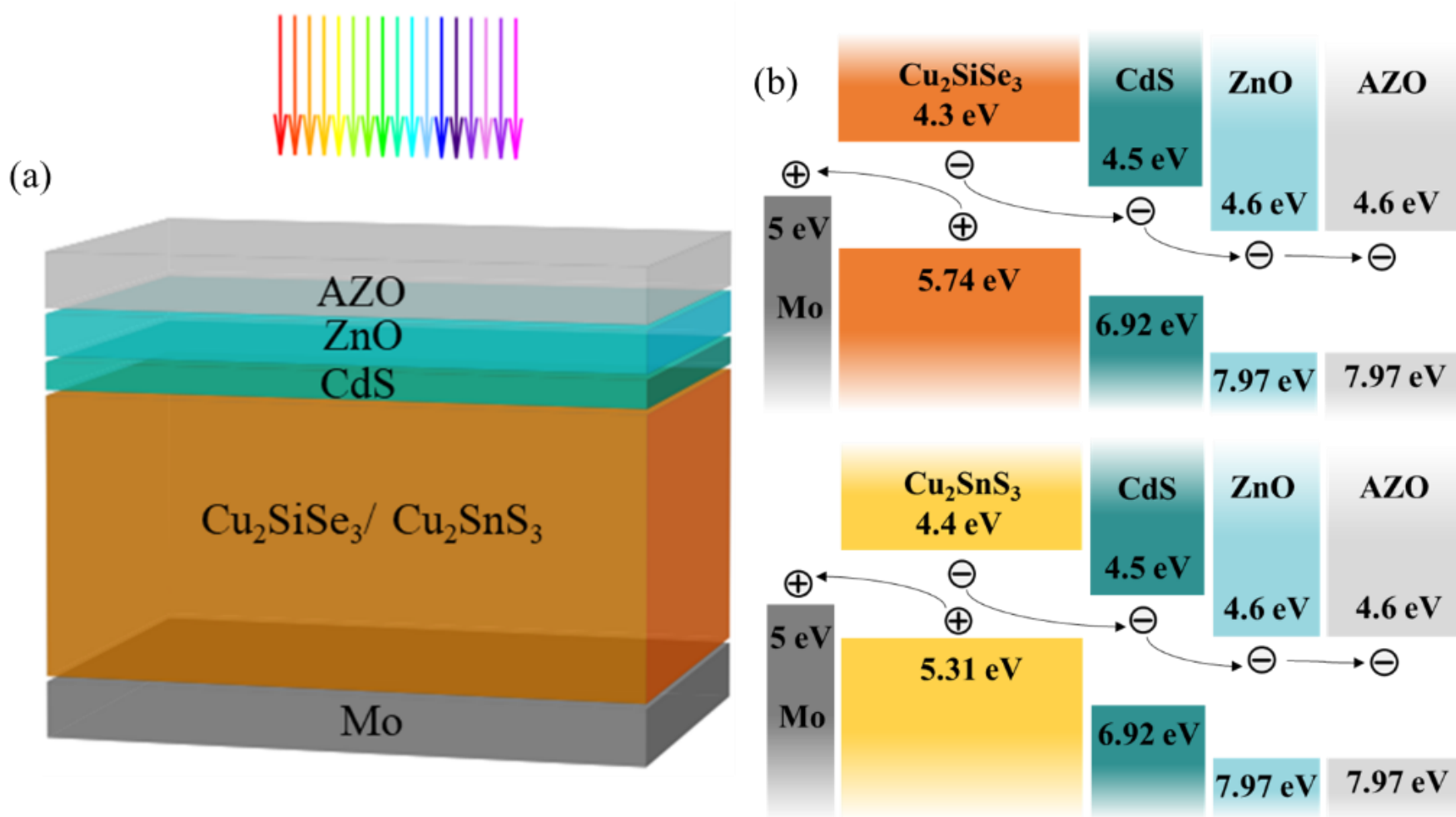

***Figure 1**: (a) Schematic of the studied $Cu_2SiSe_3$ and $Cu_2SnS_3$ absorber-based single-junction solar cell structure, (b) band offset of different layers of the $Cu_2SiSe_3$ and $Cu_2SiSe_3$ solar cell.*

## 3. Results and discussion

### 3.1 Impact of the thickness of the absorber layer

In thin-film photovoltaic devices, the absorber layer plays a pivotal role in harvesting incident solar radiation. The extent of light absorption is determined by the layer's thickness and intrinsic absorption coefficient. While increasing the absorber thickness generally enhances photon absorption and carrier generation, excessive thickness can lead to elevated recombination rates, thereby degrading overall device performance [23,24]. To investigate the influence of absorber thickness on photovoltaic characteristics, the thickness of both $Cu_2SiSe_3$ and $Cu_2SnS_3$ absorber layers was systematically varied from 0.5 µm to 4 µm. As depicted in **Figures 2a** and **2d**, both the open-circuit voltage ($V_{OC}$) and short-circuit current density ($J_{SC}$) were positively correlated with absorber thickness, attributed to increased photocarrier generation with greater light absorption. Consequently, the power conversion efficiency (PCE) also improved with increasing thickness. For $Cu_2SiSe_3$-based devices, the fill factor (FF) initially increases with thickness and reaches a saturation point at around 2 µm, whereas $V_{OC}$, $J_{SC}$, and efficiency continue to increase. In contrast, all performance parameters of the $Cu_2SnS_3$-based device increase initially but plateau beyond a thickness of approximately 2 µm. Therefore, an absorber thickness of 2 µm was identified as the optimal value for both materials, balancing enhanced absorption with minimal recombination losses.

The ratio of the charge carriers collected to the number of incident photons, also known as the external quantum efficiency (EQE), was evaluated as a function of absorber thickness for both materials, with results presented in **Figures 2b** and **2e.** These trends are consistent with the Beer-Lambert law, expressed as $I = I_0 e^{-\alpha x}$ [25], where α is the absorption coefficient, and x is the absorber thickness. The $Cu_2SiSe_3$ device exhibits efficient absorption primarily within the 400-800 nm spectral range, corresponding to the visible region, while $Cu_2SnS_3$, due to its narrower band gap, demonstrates effective absorption extending into the near-infrared region (400-1000 nm). Although thicker absorber layers lead to greater photon absorption, they also increase the likelihood of bulk recombination, particularly in low minority-carrier-lifetime materials.

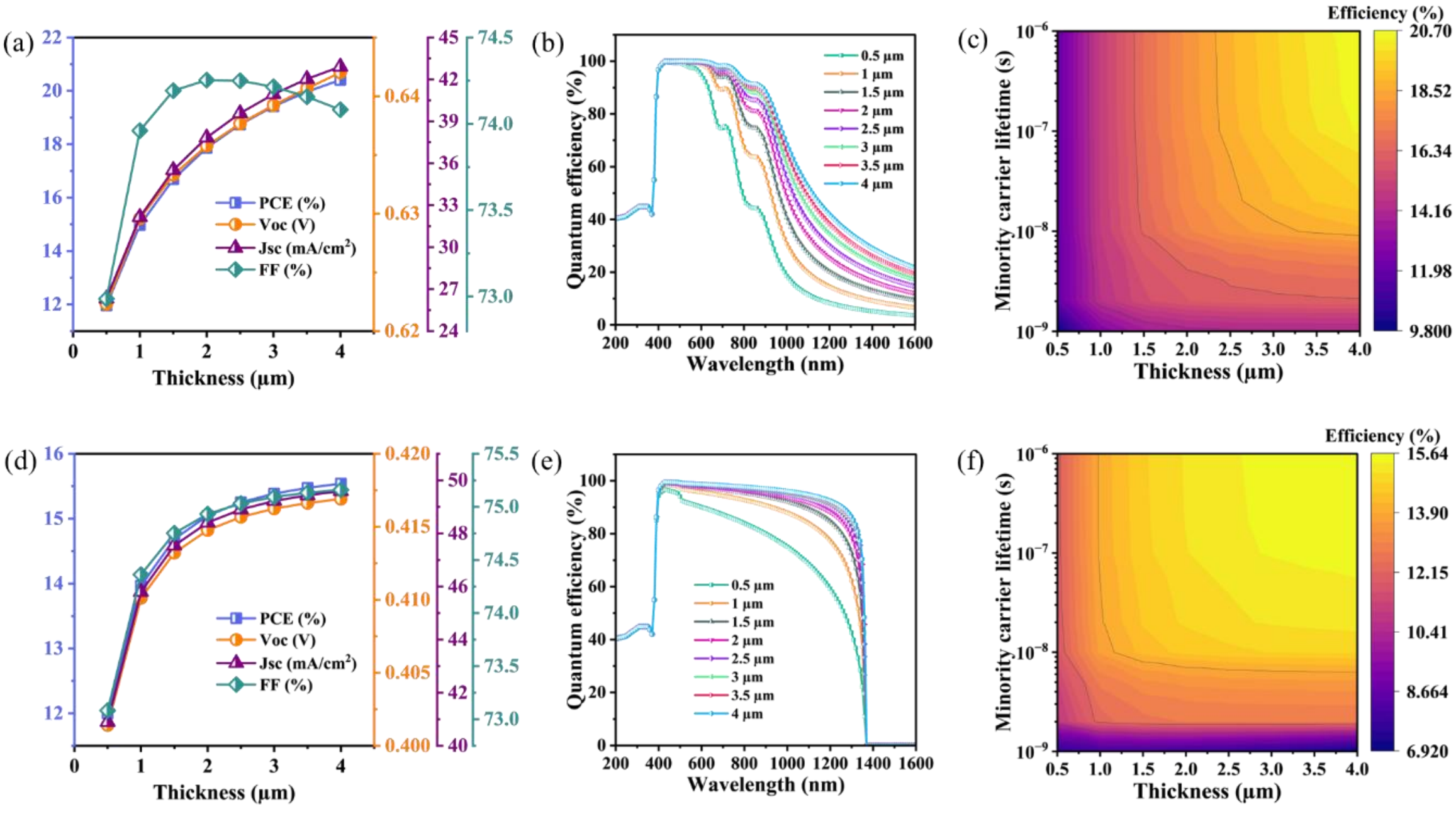

***Figure 2****: (a) and (d) Variation in efficiency, fill factor, open circuit voltage, and short circuit voltage, with absorber thickness of $Cu_2SiSe_3$ and $Cu_2SnS_3$, respectively. (b) and (e) External*

*quantum efficiency at different wavelengths at different thicknesses of $Cu_2SiSe_3$ and $Cu_2SnS_3$, respectively. (c) and (f) contour plot of efficiency for different thickness and minority carrier lifetime of the absorber layer $Cu_2SiSe_3$ and $Cu_2SnS_3$, respectively.*

The contours in **Figures 2c** and **2f** show the variation in the absorber layer thickness with the minority charge-carrier lifetime for $Cu_2SiSe_3$ and $Cu_2SnS_3$ solar cells, respectively. In both the solar cell for low minority charge-carriers and above a thickness of 1µm, the efficiency is very low and almost independent of thickness, since the excess charge carriers generated in the p-type quasi-neutral region recombine before they can reach the depletion region. The efficiency increases with increasing minority-carrier lifetime, since carriers are collected at their respective electrodes before recombining. The remaining contour plots for $Cu_2SiSe_3$ and $Cu_2SnS_3$ solar cells, showing the variation in $V_{OC}$, $J_{SC}$ and FF are presented in **Figure S1** of the supplementary material. **Figure S1a**, **S1b,** and **S1c** show the variation of $V_{OC}$, $J_{SC}$, and FF in $Cu_2SiSe_3$ solar cell. $J_{SC}$ and FF show higher values at larger minority carrier lifetime. **Figure S1d**, **S1e,** and **S1f** show the variation in $V_{OC}$, $J_{SC}$ and FF of $Cu_2SnS_3$ solar cell, and all these show a similar trend, for higher carrier lifetime and larger thickness. By improving the crystalline quality of the absorber layer, we enhance the minority-carrier lifetime[26]. Hence, in both $Cu_2SiSe_3$ and $Cu_2SnS_3$ solar cells, improving crystalline quality can achieve better results.

## 3.2 Impact of carrier and bulk defect concentration of absorber layer

The performance of thin-film solar cells is significantly influenced by the carrier concentration within the absorber layer, as it directly impacts recombination dynamics and the built-in electric field [27]. In this investigation, the carrier concentration was systematically varied from $10^{12}$ to $10^{18}$ cm$^{-3}$ for both $Cu_2SiSe_3$ and $Cu_2SnS_3$ absorber layers, results summarized in **Figures 3a** and **3c**, respectively. The $J_{SC}$ shows a significant increase for both absorber materials between $10^{14}$ and $10^{15}$ cm$^{-3}$, followed by a decline thereafter. This can be explained by the fact that, as the carrier concentration rises, more photogenerated charge carriers are produced; however, above a concentration of $10^{15}$ cm$^{-3}$, the depletion region width decreases, which will also lower the $J_{SC}$. The open-circuit voltage ($V_{OC}$) initially increases with carrier concentration. This can be explained by the dependence of $V_{OC}$ on the reverse saturation current density ($J_0$), which decreases with increasing doping, as described by the diode equation 5 [28]:

$$V_{OC} = \frac{nkT}{q} \ln\left(\frac{J_{ph}}{J_0} + 1\right) \tag{5}$$

where k is Boltzmann's constant, $J_{ph}$ is the photogenerated current density, T is the temperature. q and n are the elementary charge and diode ideality factor, respectively. The fill factor of the $Cu_2SiSe_3$ solar cell shows a trend similar to that of the $V_{OC}$, but with a sudden drop at a concentration of $5\times10^{14}$ cm$^{-3}$. The drop can be explained as a corresponding rise in the $J_{SC}$ at a constant output power. **Figure S2(a)** and **S2(b)** of the supplementary material shows the variation in the $V_{MPP}$ and $J_{MPP}$ for both solar cells. As observed, the $J_{MPP}$ increases with increasing carrier concentration, similar to the $J_{SC}$ and $V_{MPP}$, which show the same behavior as the $V_{OC}$. It all results in a stable maximum power point up to $5\times10^{16}$ cm$^{-3}$. The inverse relation of FF with the Jsc in the expression (6)

$$FF = \frac{J_{MPP}V_{MPP}}{J_{SC}V_{OC}} \quad (6)$$

causes the decline in FF at the carrier concentration of $5\times10^{14}$ cm$^{-3}$. A similar trend for FF is observed for the $Cu_2SiSe_3$ solar cell. The optimal carrier concentration for both materials is set to $5\times10^{16}$ cm$^{-3}$. At this doping level, $Cu_2SiSe_3$ achieves a peak efficiency of over 18%, while $Cu_2SnS_3$ reaches 15.59%. However, a further increase in carrier concentration led to a degradation in performance due to a reduction in the depletion region width [29,30].

In addition to doping, the concentration of bulk defects in the absorber material plays a critical role in determining minority-carrier lifetimes and thereby influencing recombination losses. The carrier lifetime ($\tau$) is inversely related to the bulk defect density ($N_t$) through the relation $\tau = 1/\sigma v_{th} N_t$, where $v_{th}$ is the thermal velocity of the charge carriers, and σ denotes the defect capture cross-section [19]. The bulk defect densities were varied from $10^{12}$ to $10^{18}$ cm$^{-3}$ for both $Cu_2SiSe_3$ and $Cu_2SnS_3$ absorber layers, and their effects on device metrics are shown in **Figures 3(b)** and **3(d)**. For $Cu_2SiSe_3$, all key performance parameters $J_{SC}$, $V_{OC}$, FF, and PCE remain relatively stable up to a defect concentration of approximately $1\times10^{17}$ cm$^{-3}$, beyond which a sharp decline is observed. This decline is attributed to the increased Shockley-Read-Hall (SRH) recombination. A similar degradation trend is also observed for $Cu_2SnS_3$. Again, as the defect concentration reaches $10^{17}$ cm$^{-3}$ or higher, SRH recombination starts to dominate, resulting in a sharp decline in photovoltaic parameters [31]. The optimized defect density is $10^{15}$ cm$^{-3}$, a realistic value for this family of chalcogenides.

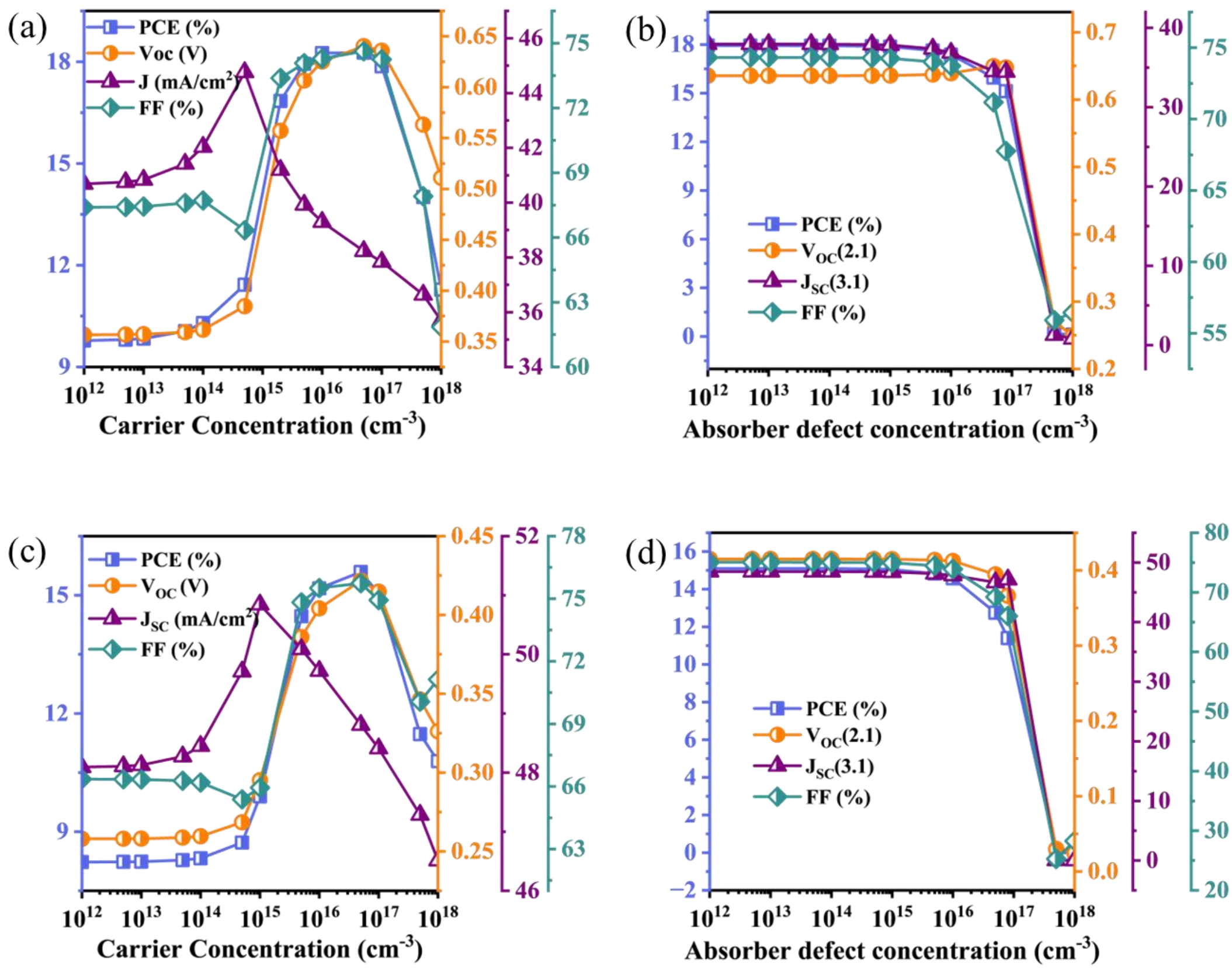

***Figure 3**: Variation in efficiency, fill factor, open circuit voltage, and short circuit current with, (a) and (c) carrier concentration of $Cu_2SiSe_3$ and $Cu_2SnS_3$, respectively, (b) and (d) defect concentration, in $Cu_2SiSe_3$ and $Cu_2SnS_3$, respectively.*

## 3.3 Impact of defect at absorber/CdS junction

Interface defects at the absorber/CdS junction primarily originate from lattice mismatch, chemical incompatibility, and constraints inherent in fabrication processes. These factors lead to structural and chemical imperfections, including dislocations, dangling bonds, and voids. The density of these interface defects ($N_d$) has a direct impact on the interface recombination velocity (S), as governed by the relation $S = \sigma N_d v_{th}$ , where $v_{th}$ is the thermal velocity, and σ denotes the capture cross-section of the charge carriers. Through Shockley-Read-Hall (SRH) recombination mechanisms, elevated interface defect densities accelerate non-radiative recombination, which severely reduces the solar cell's overall performance [32,33]. The interface defect density is varied from $10^9$ cm$^{-2}$ to $10^{16}$ cm$^{-2}$, and the results are summarised in **Figures 4(a)** and **4(d)** for $Cu_2SiSe_3$ and $Cu_2SnS_3$ solar cells, respectively. As defect density increases, key metrics such as efficiency, $V_{OC}$, $J_{SC}$, and FF decrease significantly. This degradation is attributed to increased recombination at the interface. **Figures 4(b-f)** show the effects of minority carrier lifetime and interface recombination speed on $V_{OC}$ and efficiency, with better performance for higher minority carrier lifetime and lower interface recombination speed.

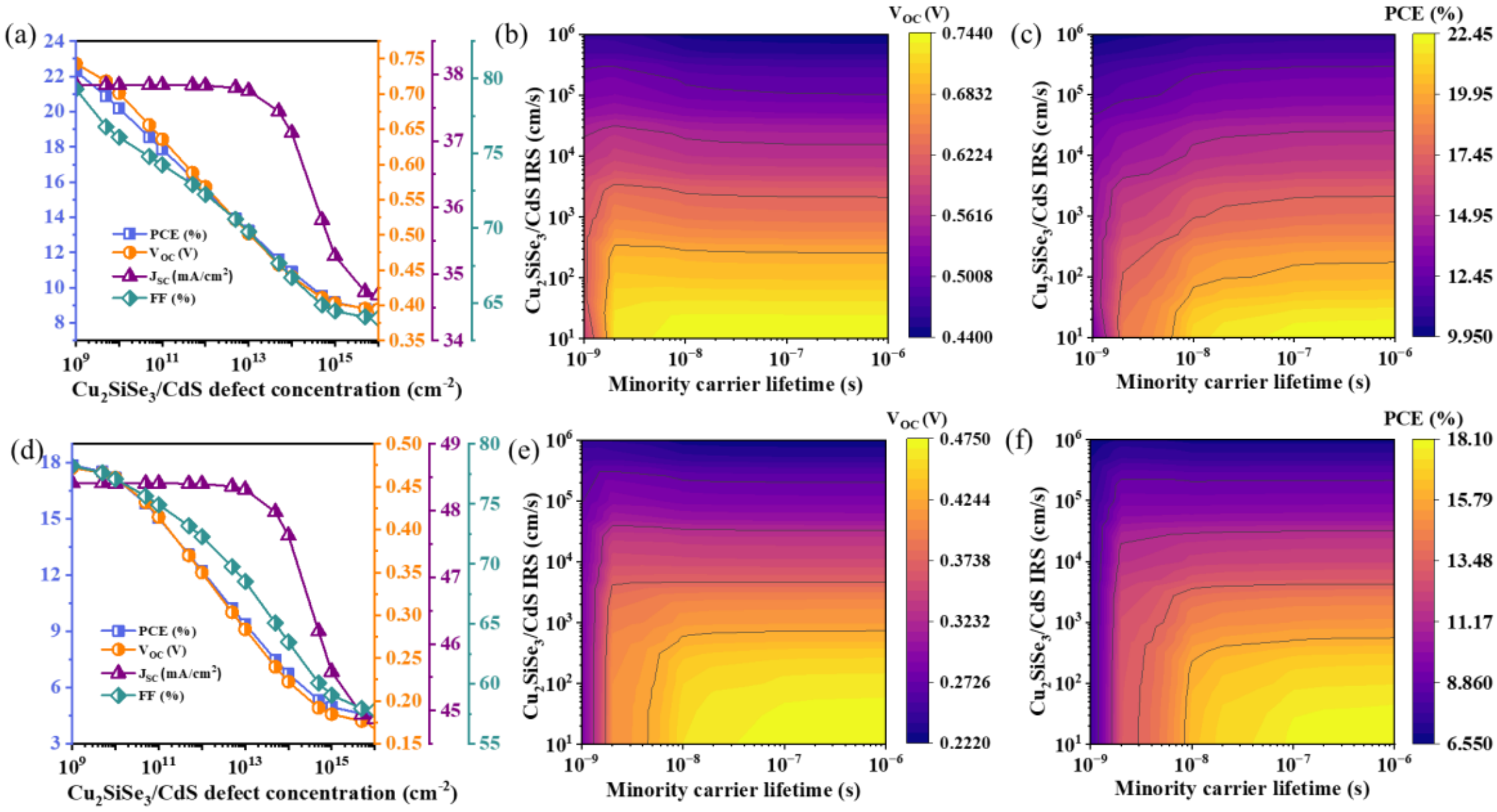

***Figure 4**: (a) and (d) Variation in efficiency fill factor, open circuit voltage, and short circuit current with interface defect concentration of $Cu_2SiSe_3$ and $Cu_2SnS_3$, respectively, (b) and (e) contour plot of the variation of open circuit voltage with minority carrier lifetime and absorber/buffer interface recombination speed (IRS) of $Cu_2SiSe_3$ and $Cu_2SnS_3$, respectively. (c) and (f) variation of efficiency with the minority carrier lifetime and absorber/buffer interface recombination speed (IRS) of $Cu_2SiSe_3$ and $Cu_2SnS_3$, respectively.*

The minority carrier lifetime is varied from $10^{-9}$ s to $10^{-6}$ s, and the interface recombination speed is varied from 10 cm/s to $10^6$ cm/s. It is found that at higher carrier lifetimes, both $V_{OC}$ and efficiency are higher. A short minority-carrier lifetime causes photogenerated carriers to recombine in the p-type quasi-neutral region before they can reach the depletion region. After a minority-carrier lifetime of ~$10^{-8}$ s, both $V_{OC}$ and efficiency are independent of the carrier lifetime and depend only on the interface recombination speed. At higher carrier lifetimes, performance increases as recombination speed decreases, because charge carriers recombine before they can reach their respective electrodes. A similar contour is plotted for the $Cu_2SnS_3$ solar cell in **Figures 4(e)** and **4(f)**. Both $V_{OC}$ and efficiency show a similar trend: for low interface recombination speed and higher minority carrier lifetime, both have maximum values. The remaining variations for the $J_{SC}$ and FF are shown in **Figure S3** of the supplementary material. **Figures S3(a)** and **S3(b)** show that for the $Cu_2SiSe_3$ solar cell, $J_{SC}$ is highest at a minority carrier-lifetime of $10^{-6}$ s and is almost independent of the interface recombination speed. The highest fill factor is obtained at higher minority-carrier lifetimes and lower interface recombination speeds. **Figures S3(c)** and **S3(d)** show the same for the $Cu_2SnS_3$ solar cell; again, the maximum value for the FF is obtained at higher minority carrier lifetime and lower interface recombination speed, and $J_{SC}$ only varies in the range of 45 $mA/cm^2$ to 49 $mA/cm^2$ and depends on the minority carrier lifetime. Therefore, reducing interface defect density through passivation or material optimization is critical to lowering recombination speed and enhancing solar cell performance. We considered an interface defect density of $10^{-11}$ $cm^{-3}$ to make the devices more realistic for these devices.

## 3.4 Impact of buffer layer

The buffer layer is essential in determining the performance of heterojunction solar cells, primarily by facilitating optimal band alignment and minimizing lattice mismatch between the absorber and window layers. A key parameter influencing device performance is the conduction-band offset (CBO) at the absorber-buffer interface, which significantly affects charge-carrier transport and recombination dynamics, thereby impacting overall efficiency. **Figures 5(a)** and **5(d)** show the dependence of solar cell performance metrics on the buffer-layer electron affinity for $Cu_2SiSe_3$ and $Cu_2SnS_3$ solar cells, respectively. The corresponding conduction-band offsets as a function of the buffer-layer CdS, electron affinity are shown in **Figures 5(b)** and **5(e)**. When the electron affinity of the buffer layer is low, a spike-type band alignment forms at the interface, indicating a positive CBO. Conversely, at higher electron affinities, the cliff forms at the interface, indicating a negative CBO [34]. According to reports [35,36], a moderate positive CBO (i.e., a spike within the range of 0 to 0.4 eV) is advantageous for device performance, as it suppresses interface recombination by creating a barrier to electron flow. In contrast, a negative CBO (cliff-type) facilitates recombination, adversely affecting device efficiency.

For the $Cu_2SiSe_3$ solar cell, optimal performance is achieved when the buffer layer's electron affinity lies between 4.1 and 4.4 eV, while the $Cu_2SnS_3$ cell performs best within the 4.2 to 4.4 eV range. The open-circuit voltage is higher at lower electron affinities due to a spike, whereas increasing the electron affinity, and thus the formation of a cliff, reduces the $V_{OC}$. Similarly, the $J_{SC}$ decreases progressively with increasing electron affinity. Consequently, higher overall efficiency is achieved at lower buffer-layer electron affinities, underscoring the importance of carefully selecting and tuning buffer-layer materials to ensure favorable band alignment and maximize photovoltaic performance.

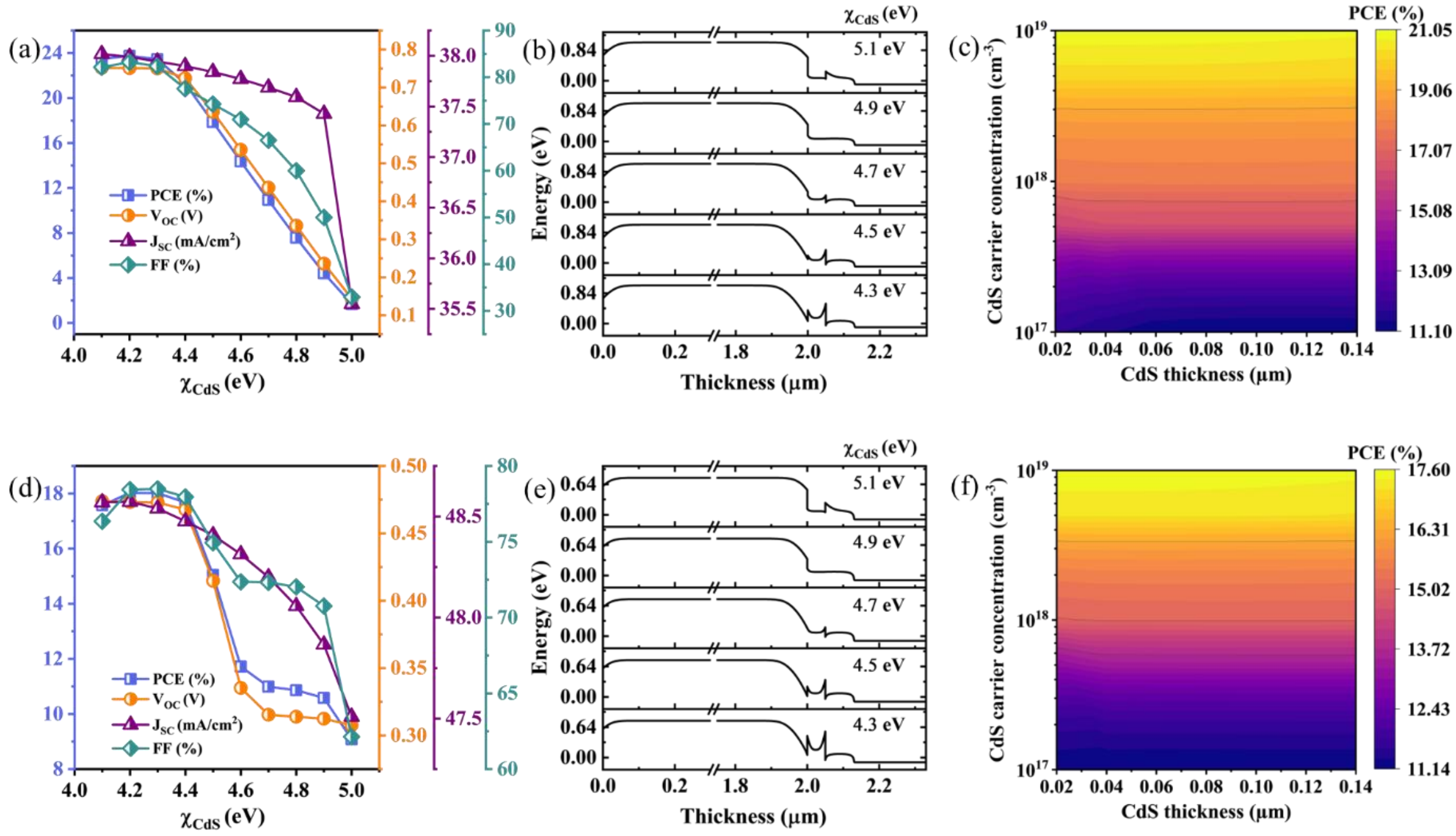

***Figure 5**: (a) and (d) Variation in efficiency, fill factor, open circuit voltage, and short circuit current with electron affinity of the buffer layer in $Cu_2SiSe_3$ and $Cu_2SnS_3$ based cells, respectively (b) and (e) conduction band alignment for different value of electron affinity of buffer layer in $Cu_2SiSe_3$ and $Cu_2SnS_3$ based cells, respectively (c) and (f) contour plot showing the variation of the efficiency with CdS carrier concentration and CdS thickness in $Cu_2SiSe_3$ and $Cu_2SnS_3$ based cells, respectively.*

To know the effect of the thickness and carrier concentration of the CdS (with acceptor concentration of the absorber layer kept fixed at $10^{16}$ $cm^{-3}$), a contour is plotted and shown in **Figures 5(c)** and **5(f)**. This shows that the PCE of both the $Cu_2SiSe_3$ and $Cu_2SnS_3$ solar cells is independent of CdS thickness, but it substantially increases as the carrier concentration increases, as the depletion width in the absorber layer increases with increasing carrier concentration in CdS. The generated electron-hole pairs can be extracted more efficiently. The variations for other parameters with thickness and carrier concentration of the CdS are shown in **Figure S4** of the supplementary material for both cells. **Figures S4(a)** and **S4(c)** show $V_{OC}$ and FF for $Cu_2SiSe_3$ solar cell, exhibiting no dependence on CdS thickness and only depend on CdS carrier concentration. The same trend is shown in **Figures S4(d)** and **S4(f)** for $Cu_2SnS_3$ solar cell. In contrast, $J_{SC}$ is maximum at lower thickness and higher carrier concentration in both cells, as shown in **Figures S4(b)** and **S4(e)**. Further, the variations in device parameters with electron affinity and interface recombination speed for both $Cu_2SiSe_3$ and $Cu_2SnS_3$ solar cells are shown in **Figures S5** and **S6** in the supplementary material. Both $V_{OC}$ and FF follow the same trend as PCE for the $Cu_2SiSe_3$ solar cell, with higher values at a lower interface recombination speed. Further, $J_{SC}$ is almost unaffected by CdS electron affinity over the entire range of interface recombination speed, with very small variation at very high CdS electron affinity and low interface recombination speed. For the $Cu_2SnS_3$ solar cell, again, the best PCE, FF, and $V_{OC}$ are obtained at a lower interface recombination speed and an electron affinity of ~ 4.3 eV. In contrast, the $J_{SC}$ only varied with the electron affinity of CdS.

## 3.5 Impact of the work function of the back contact

We have used Mo with a work function of 5 eV, which is commonly used in quaternary and ternary chalcogenide solar cells. We have investigated the effect of the back-contact work function by varying it from 4.7 eV to 5.6 eV for Cu2SiSe3 and from 4.4 eV to 5.3 eV for Cu2SnS3 solar cells, and the results are shown in **Figures 6(a) and 6(d),** respectively. As the work function of the metal back contact increases, all photovoltaic parameters improve, as a higher work function results in an Ohmic contact, facilitating the extraction of the majority charge carriers. In contrast, a lower work function results in a Schottky barrier at the absorber/metal interface. [37] For $Cu_2SiSe_3$ solar cell, $V_{OC}$, FF, and efficiency initially increase, and saturate beyond 5.2 eV workfunction. Initially, $J_{SC}$ increases slowly, then rises rapidly above 5.5 eV. Similarly, for the $Cu_2SnS_3$ solar cell, $V_{OC}$, FF, and PCE increase with the work function until 4.8 eV, then saturate, whereas $J_{SC}$ increased abruptly after 4.4eV and then saturated. We have also shown the band diagrams of the device for the work functions 5, 5.3, and 5.6 eV in **Figure 6(b)** for $Cu_2SiSe_3$, and **Figure 6(e)** shows the same for $Cu_2SnS_3$ solar cell, for different work functions of ~ 4.7, 5.0, and 5.3 eV. The barrier height increases as the work function increases due to band bending at the absorber/back contact interface, which effectively blocks electrons and only extracts holes, thereby reducing recombination.

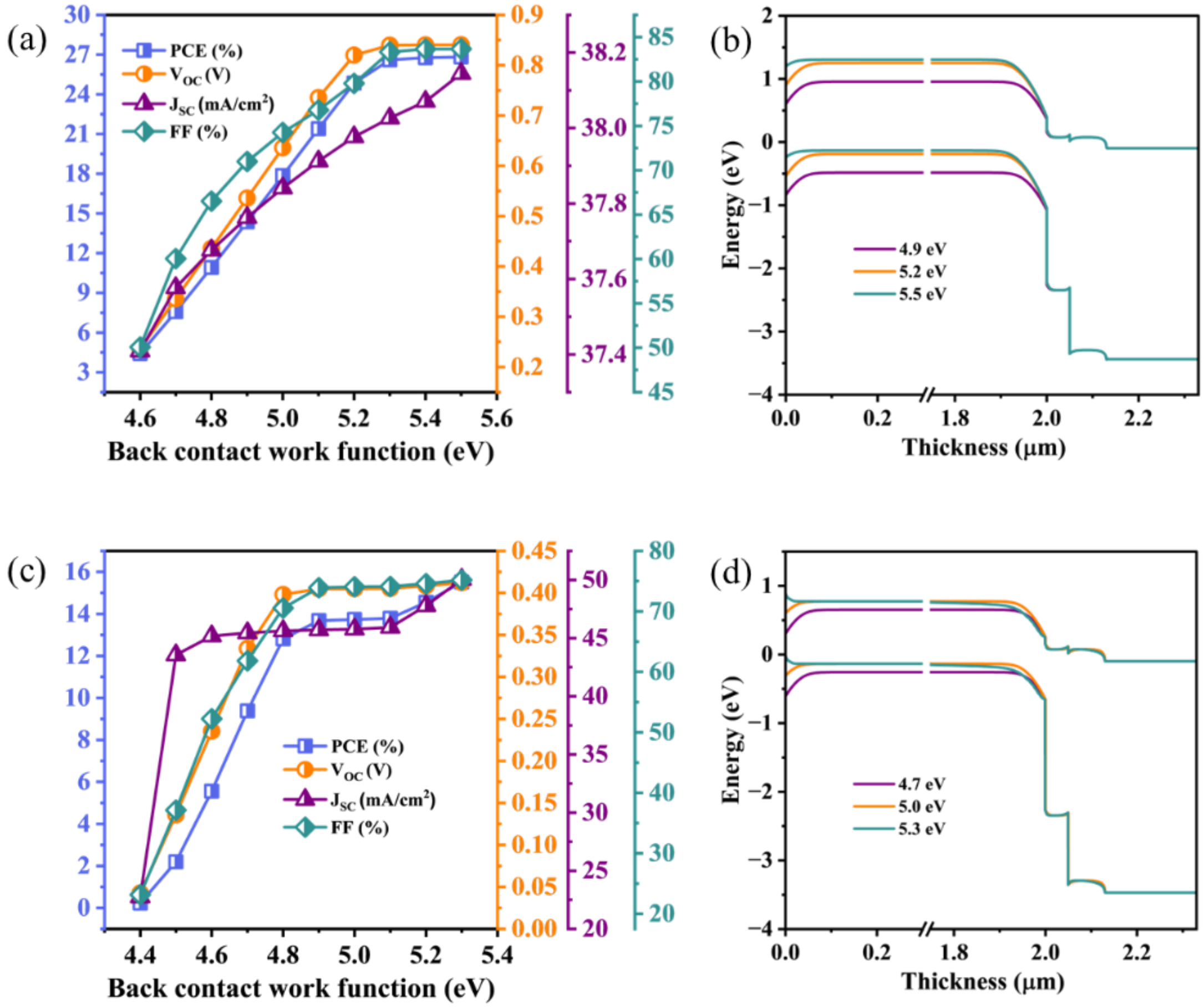

***Figure 6**: (a) and (c) Variation in efficiency, fill factor, open circuit voltage, and short circuit current with work function of back contact in $Cu_2SiSe_3$ and $Cu_2SnS_3$-based cells, respectively, (b) and (d) band diagram of the proposed single junction solar cells based on $Cu_2SiSe_3$ and $Cu_2SnS_3$, respectively, for different work functions of the back contact*

## 3.6 Tandem solar cell of $Cu_2SiSe_3/Cu_2SnS_3$

Single-junction solar cells are inherently limited in their ability to utilize the full spectrum of incident solar radiation. They can only absorb photons with energies greater than the absorber layer's band gap, while lower-energy photons pass through without being absorbed. To harness this unutilized portion of the spectrum, tandem solar cell architectures are employed, in which two or more absorbers with dissimilar band gaps are stacked. In such a configuration, the top cell with a higher band gap absorbs high-energy photons, while the bottom cell with a narrower band gap absorbs low-energy photons transmitted through the top cell.

Among the various tandem configurations, the two most prominent are the two-terminal (2T) and four-terminal (4T) designs. The 4T configuration offers fabrication simplicity and eliminates the need for current matching between sub-cells. However, it suffers from parasitic optical losses due to reflections and absorptions at the intermediate interface, which generally requires more materials, thereby increasing manufacturing costs [38,39]. The theoretical efficiency limit of a dual-junction tandem solar cell under ideal conditions is approximately 46%, with optimal band gap combinations of 0.96 eV and 1.65 eV for the 2T configuration, and 0.95 eV and 1.74 eV for the 4T configuration [39,40]. It is important to note that these theoretical limits assume idealized conditions, including perfect absorption, unity quantum efficiency, no recombination losses, and ideal diode behaviour [41]. While achieving such performance in practice is challenging due to various non-idealities and material limitations, these theoretical models remain invaluable in guiding the design and optimization of high-efficiency tandem devices.

We focused on simulating a 2T tandem solar cell, schematically illustrated in **Figure 7a**. **Figure 7b** represents the flow of the charge carriers and recombination of holes from the top cell and electrons from the bottom cell at the $Cu_2SiSe_3$/AZO interface. The proposed architecture features a $Cu_2SiSe_3$ top cell with a band gap of 1.44 eV and a $Cu_2SnS_3$ bottom cell with a band gap of 0.91 eV. The recombination layer that connects the two sub-cells is a key component of the 2T tandem configuration. This layer plays a critical role in preventing charge accumulation at the interface. It must exhibit high transparency and electrical conductivity to facilitate efficient recombination of electrons from the bottom cell with holes from the top cell. Commonly used materials for recombination layers include molybdenum-doped indium oxide (IMO), indium tin oxide (ITO), fluorine-doped tin oxide (FTO), and aluminium-doped zinc oxide (AZO) [42–44]. In the proposed structure, AZO, which originally served as the window layer of the bottom cell, also functions as the recombination layer.

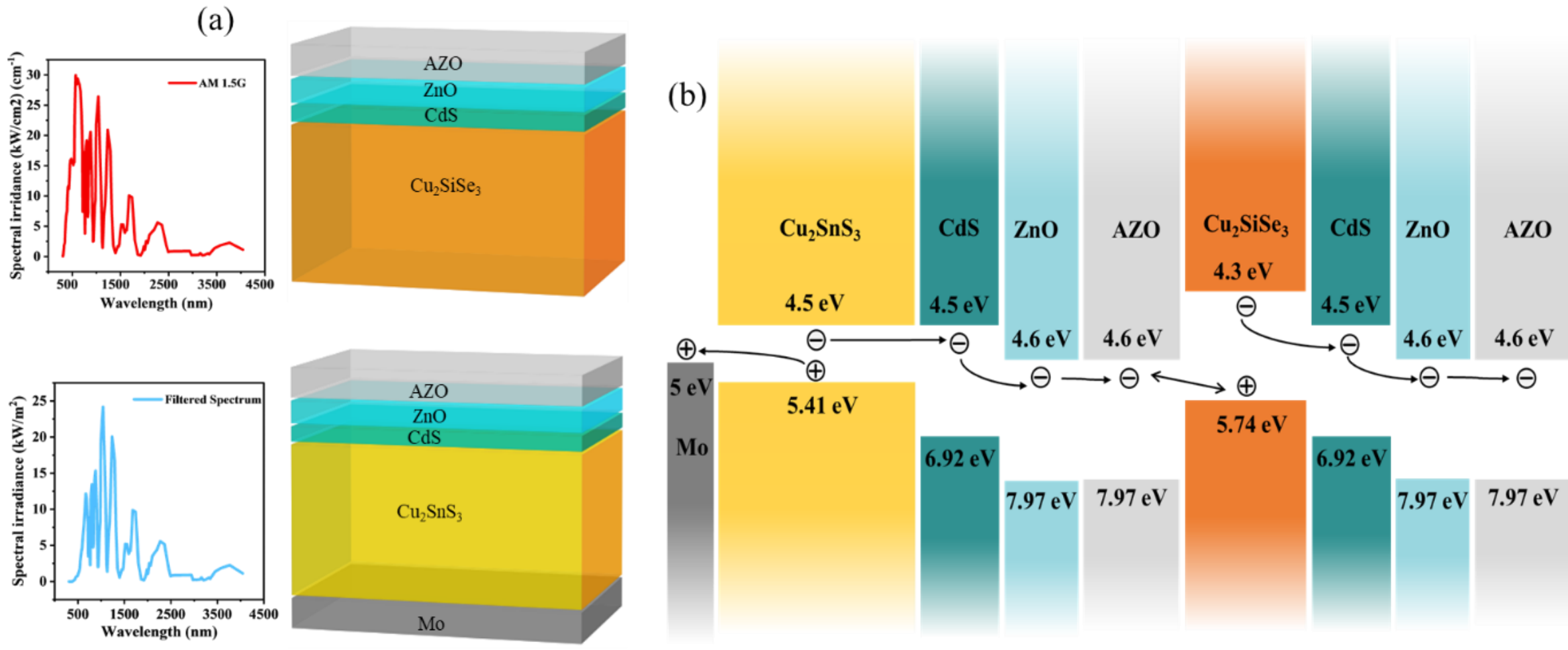

***Figure** 7: (a) Schematic diagram of the tandem solar cell, (b) energy offsets of different layers of the tandem solar cell.*

We followed the process discussed in Section 3 of the methodology to determine the current matching condition. **Figure S7** in the supplementary material presents a contour plot of the difference in $J_{SC}$ between the two sub-cells across various combinations of top and bottom absorber thicknesses. Notably, as the bottom cell thickness increases from 0.5 µm to 2.5 µm, the corresponding top cell thickness required for current matching varies only modestly, from 0.15 µm to 0.21 µm.

**Figure 8a** shows the $J_{SC}$ of both sub-cells as a function of the $Cu_2SiSe_3$ top cell thickness. As the top absorber thickness increases, $J_{SC}$ in the top cell rises due to enhanced light absorption and photogeneration. However, the bottom cell $J_{SC}$ decreases with a fixed bottom absorber thickness, as a thicker top layer filters out more incident light, reducing the photon flux reaching the lower cell. Similarly, increasing the thickness of the $Cu_2SnS_3$ layer increases $J_{SC}$ due to enhanced light absorption. The intersection points of the $J_{SC}$ curves represent the current-matching conditions required for optimal 2T operation.

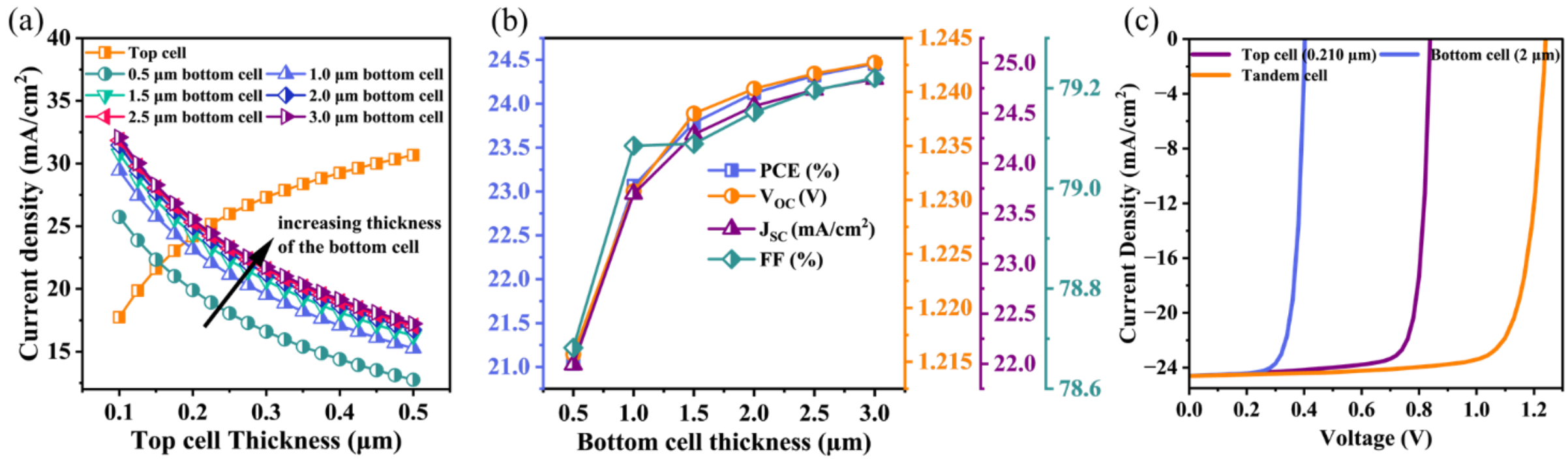

***Figure 8**: (a) Variation of $J_{SC}$ of $Cu_2SiSe_3$ top sub cell and $Cu_2SnS_3$ bottom sub cell at different thicknesses of the top cell (b) Variation of $V_{OC}$, $J_{SC}$, PCE, and FF as a function of bottom cell thickness. (c) I-V characteristic under matched current conditions of the top, bottom and tandem solar cells.*

The tandem cell's key performance metrics, $V_{OC}$, $J_{SC}$, FF, and PCE, are plotted in **Figure 8b**, as functions of bottom cell thickness. As the $Cu_2SnS_3$ layer becomes thicker, overall efficiency improves, but both efficiency and FF saturate beyond 2 µm. The optimal configuration is achieved with 0.210 µm for the $Cu_2SiSe_3$ top absorber and 2.1 µm for the $Cu_2SnS_3$ bottom absorber, resulting in current matching. Under this configuration, the tandem device achieves a $V_{OC}$ of 1.24 V, $J_{SC}$ of 24.6 mA/cm², FF of 79.2%, and overall efficiency of 24.1%, as shown in the I-V characteristics in **Figure 8c**.

Compared with other experimentally reported tandem solar cells, particularly those based on perovskite and inorganic thin-film technologies, the $Cu_2SiSe_3$/$Cu_2SnS_3$ tandem design exhibits competitive performance. For example, a perovskite/silicon tandem has demonstrated a record efficiency of 32.5%, with an open-circuit voltage of 1.97 V and a short-circuit current density of 20.24 mA/cm² [45]. All-perovskite tandem has achieved 29.7% efficiency, with $V_{OC}$, $J_{SC}$, and FF values of 2.17 V, 16.4 mA/cm², and 83.3%, respectively [46]. The champion perovskite/CIGS tandem device reported [47], achieving 24.6% efficiency, with an open-circuit voltage of 1.7 V, a short-circuit current density of 20.97 mA/cm², and an FF of 67.3%, comparable to the proposed $Cu_2SiSe_3$/$Cu_2SnS_3$ tandem structure. The achieved $J_{SC}$ of 24.6 mA/cm² and FF of 79.2% are particularly noteworthy, indicating effective carrier collection and low series resistance. The high FF substantially contributes to boosting the device's overall efficiency.

## 4. Conclusion

In this study, we have conducted a comprehensive simulation of single-junction solar cells employing $Cu_2SiSe_3$ and $Cu_2SnS_3$ as absorber materials. Both absorber materials, $Cu_2SiSe_3$ with a band gap of 1.44 eV and $Cu_2SnS_3$ with a band gap of 0.91 eV, were optimized individually by systematically varying key parameters, including absorber thickness, defect density, and carrier concentration, in both the bulk and at interfaces, under standard AM1.5G solar illumination. The optimized $Cu_2SiSe_3$-based single junction cell achieved an efficiency of 18.13%, with a $V_{OC}$ of 0.64 V, $J_{SC}$ of 38 mA/cm², and FF of 74.62%. Similarly, the optimized $Cu_2SnS_3$ cell yielded $V_{OC}$, $J_{SC}$, FF, and PCE of 0.42 V, 48.80 mA/cm², 75.74%, and 15.59%, respectively. The influence of the buffer layer was also analyzed, with particular focus on the electron affinity and the resulting conduction band alignment. The noticed spike-type conduction-band offset at the buffer/absorber interface is more favorable than a cliff-type alignment, which can lead to higher recombination losses. The increase in back-contact's work function led to improved photovoltaic parameters, attributed to enhanced band bending at the absorber/contact interface. Further, a 2T tandem solar cell architecture employing $Cu_2SiSe_3$ and $Cu_2SnS_3$ as absorbers in the top and bottom cells, respectively, showed an overall $V_{OC}$ of 1.24 V, $J_{SC}$ of 24.6 mA/cm², FF of 79.2%, and a PCE of 24.1%, substantially outperforming the standalone single-junction devices. This work demonstrates the potential of combining $Cu_2SiSe_3$ and $Cu_2SnS_3$ in a tandem configuration. It provides a means to further optimize device performance by tailoring the band gaps of these materials with suitable dopants. The results also contribute to ongoing efforts to develop non-toxic, earth-abundant, and cost-effective alternatives to conventional photovoltaic technologies, advancing the pursuit of sustainable, environmentally friendly energy solutions.

## 5. Supplementary Information

The supplementary file includes Materials parameters of each layer used for the simulation, Contour plots for photovoltaic parameters comparing the thickness and minority carrier lifetime for both $Cu_2SiSe_3$ and $Cu_2SnS_3$ based cell, variation in the $V_{MPP}$ and $J_{MPP}$ with the carrier concentration for both $Cu_2SiSe_3$ and $Cu_2SnS_3$ based cell, Contour plots for photovoltaic parameters comparing the minority carrier lifetime for both $Cu_2SiSe_3$ and $Cu_2SnS_3$ based cell, Contour plots for photovoltaic parameters comparing the thickness and carrier concentration of CdS for both $Cu_2SiSe_3$ and $Cu_2SnS_3$ based cell, Contour plots for photovoltaic parameters comparing the CdS electron affinity and $Cu_2SiSe_3$/CdS interface recombination speed, Contour plots for photovoltaic parameters comparing the CdS electron affinity and $Cu_2SnS_3$/CdS interface recombination speed.

## 6. Acknowledgment

Ambesh Dixit acknowledges ANRF (SERB)-DST, Govt. of India, for funding support through projects #CRG/2020/004023 and #SERB/F/10090/2021-2022 for this work. Authors also acknowledge the HPC facility at IIT Jodhpur for providing computing facilities. Saptarshi Mandal acknowledges A-MAD Laboratory members for providing their valuable inputs.

## Conflict of interest statement:

The authors do not have any conflicts of interest for the present work.

### References

(1) Yoshikawa, K.; Kawasaki, H.; Yoshida, W.; Irie, T.; Konishi, K.; Nakano, K.; Uto, T.; Adachi, D.; Kanematsu, M.; Uzu, H.; Yamamoto, K. Silicon Heterojunction Solar Cell with Interdigitated Back Contacts for a Photoconversion Efficiency over 26%. *Nat Energy* **2017**, *2* (5). https://doi.org/10.1038/nenergy.2017.32.

(2) Maldonado, S. The Importance of New "Sand-To-Silicon" Processes for the Rapid Future Increase of Photovoltaics. *ACS Energy Letters*. American Chemical Society November 13, 2020, pp 3628–3632. https://doi.org/10.1021/acsenergylett.0c02100.

(3) Saga, T. Advances in Crystalline Silicon Solar Cell Technology for Industrial Mass Production. *NPG Asia Materials*. July 2010, pp 96–102. https://doi.org/10.1038/asiamat.2010.82.

(4) Kayes, B. M.; Nie, H.; Twist, R.; Spruytte, S. G.; Reinhardt, F.; Kizilyalli, I. C.; Higashi, G. S. 27.6% Conversion Efficiency, a New Record for Single-Junction Solar Cells under 1 Sun Illumination. In *Conference Record of the IEEE Photovoltaic Specialists Conference*; 2011; pp 000004–000008. https://doi.org/10.1109/PVSC.2011.6185831.

(5) Green, M.; Dunlop, E.; Hohl-Ebinger, J.; Yoshita, M.; Kopidakis, N.; Hao, X. Solar Cell Efficiency Tables (Version 57). *Progress in Photovoltaics: Research and Applications* **2021**, *29*, 3–15. https://doi.org/10.1002/pip.3371.

(6) Green, M. A.; Dunlop, E. D.; Yoshita, M.; Kopidakis, N.; Bothe, K.; Siefer, G.; Hao, X.; Jiang, J. Y. Solar Cell Efficiency Tables (Version 65). *Progress in Photovoltaics: Research and Applications* **2024**. https://doi.org/10.1002/pip.3867.

(7) Nakamura, M.; Yamaguchi, K.; Kimoto, Y.; Yasaki, Y.; Kato, T.; Sugimoto, H. Cd-Free Cu(In,Ga)(Se,S)2 Thin-Film Solar Cell with Record Efficiency of 23.35%. *IEEE Journal of Photovoltaics* **2019**, *9*, 1863–1867. https://doi.org/10.1109/JPHOTOV.2019.2937218.

(8) Zhou, J.; Xu, X.; Wu, H.; Wang, J.; Lou, L.; Yin, K.; Gong, Y.; Shi, J.; Luo, Y.; Li, D.; Xin, H.; Meng, Q. Control of the Phase Evolution of Kesterite by Tuning of the Selenium Partial Pressure for Solar Cells with 13.8% Certified Efficiency. *Nature Energy* **2023**, *8*, 526–535. https://doi.org/10.1038/s41560-023-01251-6.

(9) Aldakov, D.; Lefrançois, A.; Reiss, P. Ternary and Quaternary Metal Chalcogenide Nanocrystals: Synthesis, Properties and Applications. *Journal of Materials Chemistry C* **2013**, *1*, 3756–3776. https://doi.org/10.1039/c3tc30273c.

(10) Shen, C.; Li, T.; Zhang, Y.; Xie, R.; Long, T.; Fortunato, N. M.; Liang, F.; Dai, M.; Shen, J.; Wolverton, C. M.; Zhang, H. Accelerated Screening of Ternary Chalcogenides for Potential Photovoltaic Applications. *Journal of the American Chemical Society* **2023**, *145*, 21925–21936. https://doi.org/10.1021/jacs.3c06207.

(11) Cheng, A.-J.; Manno, M.; Khare, A.; Leighton, C.; Campbell, S. A.; Aydil, E. S. Imaging and Phase Identification of Cu2ZnSnS4 Thin Films Using Confocal Raman Spectroscopy. *Journal of Vacuum Science & Technology A: Vacuum, Surfaces, and Films* **2011**, *29*. https://doi.org/10.1116/1.3625249.

(12) Siebentritt, S.; Schorr, S. Kesterites-a Challenging Material for Solar Cells. *Progress in Photovoltaics: Research and Applications* **2012**, *20*, 512–519. https://doi.org/10.1002/pip.2156.

(13) Kanai, A.; Toyonaga, K.; Chino, K.; Katagiri, H.; Araki, H. Fabrication of Cu2SnS3 Thin-Film Solar Cells with Power Conversion Efficiency of over 4%. In *Japanese Journal of Applied Physics*; Japan Society of Applied Physics, 2015; Vol. 54. https://doi.org/10.7567/JJAP.54.08KC06.

(14) Umehara, M.; Tajima, S.; Aoki, Y.; Takeda, Y.; Motohiro, T. Cu2Sn1-XGexS3 Solar Cells Fabricated with a Graded Bandgap Structure. *Applied Physics Express* **2016**, *9*. https://doi.org/10.7567/APEX.9.072301.

(15) Kanai, A.; Sugiyama, M. Na Induction Effects for J–V Properties of Cu2SnS3 (CTS) Solar Cells and Fabrication of a CTS Solar Cell over-5.2% Efficiency. *Solar Energy Materials and Solar Cells* **2021**, *231*. https://doi.org/10.1016/j.solmat.2021.111315.

(16) Kukreti, S.; Ramawat, S.; Dixit, A. Band Anisotropy and Quartic Anharmonicity Cooperate to Drive p -Type Thermoelectricity in the Ternary Diamondlike Semiconductor Cu2SiSe3. *Physical Review B* **2024**, *110*. https://doi.org/10.1103/PhysRevB.110.245202.

(17) Nicolson, A.; Kavanagh, S. R.; Savory, C. N.; Watson, G. W.; Scanlon, D. O. Cu2SiSe3 as a Promising Solar Absorber: Harnessing Cation Dissimilarity to Avoid Killer Antisites. *Journal of Materials Chemistry A* **2023**, *11* (27), 14833–14839. https://doi.org/10.1039/d3ta02429f.

(18) Mathur, A. S.; Upadhyay, S.; Singh, P. P.; Sharma, B.; Arora, P.; Rajput, V. K.; Kumar, P.; Singh, D.; Singh, B. P. Role of Defect Density in Absorber Layer of Ternary Chalcogenide Cu2SnS3 Solar Cell. *Optical Materials* **2021**, *119*, 111314. https://doi.org/10.1016/j.optmat.2021.111314.

(19) Burgelman, M.; Decock, K.; Niemegeers, A.; Verschraegen, J.; Degrave, S. *SCAPS Manual*; 2016.

(20) Ramawat, S.; Kukreti, S.; Sapkota, D. J.; Dixit, A. Insight into the Microband Offset and Charge Transport Layer's Suitability for an Efficient Inverted Perovskite Solar Cell: A Case Study for Tin-Based B-γ-CsSnI3. *Energy and Fuels* **2024**, *38*, 9011–9026. https://doi.org/10.1021/acs.energyfuels.4c00763.

(21) Ramawat, S.; Sapkota, D. J.; Kukreti, S.; Dixit, A. Prospective $CaSnS_3$/Si Tandem Solar Cells: Assessing Material Aspects and Quantifying Temperature-Induced Efficiency Losses. *J Mater Chem A Mater* **2025**, *13* (44), 38180–38193. https://doi.org/10.1039/D5TA04623H.

(22) Salah, M. M.; Saeed, A.; Mousa, M.; Abouelatta, M.; Zekry, A.; Shaker, A.; Amer, F. Z.; Mubarak, R. I. Numerical Analysis of Carbon-Based Perovskite Tandem Solar Cells: Pathways towards High Efficiency and Stability. *Renewable and Sustainable Energy Reviews* **2024**, *189*. https://doi.org/10.1016/j.rser.2023.114041.

(23) Nelson, J. Diffusion-Limited Recombination in Polymer-Fullerene Blends and Its Influence on Photocurrent Collection. *Phys Rev B Condens Matter Mater Phys* **2003**, *67*. https://doi.org/10.1103/PhysRevB.67.155209.

(24) Mouĺ, A. J.; Bonekamp, J. B.; Meerholz, K. The Effect of Active Layer Thickness and Composition on the Performance of Bulk-Heterojunction Solar Cells. *J Appl Phys* **2006**, *100*. https://doi.org/10.1063/1.2360780.

(25) Kukreti, S.; Gupta, G. K.; Dixit, A. Theoretical DFT Studies of Cu2HgSnS4 Absorber Material and Al:ZnO/ZnO/CdS/Cu2HgSnS4/Back Contact Heterojunction Solar Cell. *Solar Energy* **2021**, *225*, 802–813. https://doi.org/10.1016/j.solener.2021.07.071.

(26) Courel, M.; Andrade-Arvizu, J. A.; Vigil-Galán, O. The Role of Buffer/Kesterite Interface Recombination and Minority Carrier Lifetime on Kesterite Thin Film Solar Cells. *Mater Res Express* **2016**, *3* (9). https://doi.org/10.1088/2053-1591/3/9/095501.

(27) Kirchartz, T.; Cahen, D. Minimum Doping Densities for p–n Junctions. *Nature Energy*. Nature Research December 2020, pp 973–975. https://doi.org/10.1038/s41560-020-00708-2.

(28) Moustafa, M.; Al Zoubi, T.; Yasin, S. Exploration of CZTS-Based Solar Using the ZrS2 as a Novel Buffer Layer by SCAPS Simulation. *Optical Materials* **2022**, *124*. https://doi.org/10.1016/j.optmat.2022.112001.

(29) Reinders, A.; Verlinden, Pierre.; Sark, W. van.; Freundlich, Alexandre. *Photovoltaic Solar Energy : From Fundamentals to Applications*; John Wiley & Sons Ltd, 2017.

(30) Kukreti, S.; Sapkota, D. J.; Dixit, A. Designing Graded Band Gap Active Layer $Cu_2HgSn(S_{1-x}Se_x)_4$ over Complex Tandem Structure for Efficient Photovoltaic Cells with Efficiency >20%. *Energy & Fuels* **2023**, *37* (16), 12335–12344. https://doi.org/10.1021/acs.energyfuels.3c01257.

(31) Kowsar, A.; Shafayet-Ul-Islam, M.; Ali Shaikh, M. A.; Palash, M. L.; Kuddus, A.; Uddin, M. I.; Farhad, S. F. U. Enhanced Photoconversion Efficiency of Cu2MnSnS4 Solar Cells by Sn-/Zn-Based Oxides and Chalcogenides Buffer and Electron Transport Layers. *Solar Energy* **2023**, *265*. https://doi.org/10.1016/j.solener.2023.112096.

(32) Chen, S.; Walsh, A.; Gong, X. G.; Wei, S. H. Classification of Lattice Defects in the Kesterite Cu2ZnSnS 4 and Cu2ZnSnSe4 Earth-Abundant Solar Cell Absorbers. *Advanced Materials* **2013**, *25*, 1522–1539. https://doi.org/10.1002/adma.201203146.

(33) Zhang, X.; Zhou, Z.; Cao, L.; Kou, D.; Yuan, S.; Zheng, Z.; Yang, G.; Tian, Q.; Wu, S.; Liu, S. Suppressed Interface Defects by GeSe2 Post-Deposition Treatment Enables High-Efficiency Kesterite Solar Cells. *Adv Funct Mater* **2023**, *33*. https://doi.org/10.1002/adfm.202211315.

(34) Bagade, S. S.; Barik, S. B.; Malik, M. M.; Patel, P. K. Impact of Band Alignment at Interfaces in Perovskite-Based Solar Cell Devices. *Materials Today: Proceedings* **2023**. https://doi.org/10.1016/j.matpr.2023.02.117.

(35) Minemoto, T.; Matsui, T.; Takakura, H.; Hamakawa, Y.; Negami, T.; Hashimoto, Y.; Uenoyama, T.; Kitagawa, M. Theoretical Analysis of the Effect of Conduction Band Offset of Window/CIS Layers on Performance of CIS Solar Cells Using Device Simulation. *Solar Energy Materials and Solar Cells* **2001**, *67*, 83–88. https://doi.org/10.1016/S0927-0248(00)00266-X.

(36) Gansukh, M.; Li, Z.; Rodriguez, M. E.; Engberg, S.; Martinho, F. M. A.; Mariño, S. L.; Stamate, E.; Schou, J.; Hansen, O.; Canulescu, S. Energy Band Alignment at the Heterointerface between CdS and Ag-Alloyed CZTS. *Scientific Reports* **2020**, *10*. https://doi.org/10.1038/s41598-020-73828-0.

(37) Brillson, L. J.; Lu, Y. ZnO Schottky Barriers and Ohmic Contacts. *Journal of Applied Physics*. June 2011. https://doi.org/10.1063/1.3581173.

(38) Eperon, G. E.; Leijtens, T.; Bush, K. A.; Prasanna, R.; Green, T.; Wang, J. T. W.; McMeekin, D. P.; Volonakis, G.; Milot, R. L.; May, R.; Palmstrom, A.; Slotcavage, D. J.; Belisle, R. A.; Patel, J. B.; Parrott, E. S.; Sutton, R. J.; Ma, W.; Moghadam, F.; Conings, B.; Babayigit, A.; Boyen, H. G.; Bent, S.; Giustino, F.; Herz, L. M.; Johnston, M. B.; McGehee, M. D.; Snaith, H. J. Perovskite-Perovskite Tandem Photovoltaics with Optimized Band Gaps. *Science* **2016**, *354*, 861–865. https://doi.org/10.1126/science.aaf9717.

(39) Leijtens, T.; Bush, K. A.; Prasanna, R.; McGehee, M. D. Opportunities and Challenges for Tandem Solar Cells Using Metal Halide Perovskite Semiconductors. *Nature Energy*. Nature Publishing Group October 2018, pp 828–838. https://doi.org/10.1038/s41560-018-0190-4.

(40) Xu, L.; Xu, F.; Liu, J.; Zhang, X.; Subbiah, A. S.; De Wolf, S. Bandgap Optimization for Bifacial Tandem Solar Cells. *ACS Energy Letters* **2023**, *8*, 3114–3121. https://doi.org/10.1021/acsenergylett.3c01014.

(41) Abdul Hadi, S.; Fitzgerald, E. A.; Nayfeh, A. Theoretical Efficiency Limit for a Two-Terminal Multi-Junction "Step-Cell" Using Detailed Balance Method. *Journal of Applied Physics* **2016**, *119*. https://doi.org/10.1063/1.4942223.

(42) Albrecht, S.; Saliba, M.; Correa Baena, J. P.; Lang, F.; Kegelmann, L.; Mews, M.; Steier, L.; Abate, A.; Rappich, J.; Korte, L.; Schlatmann, R.; Nazeeruddin, M. K.; Hagfeldt, A.; Grätzel, M.; Rech, B. Monolithic Perovskite/Silicon-Heterojunction Tandem Solar Cells Processed at Low Temperature. *Energy and Environmental Science* **2016**, *9*, 81–88. https://doi.org/10.1039/c5ee02965a.

(43) Sahli, F.; Kamino, B. A.; Werner, J.; Bräuninger, M.; Paviet-Salomon, B.; Barraud, L.; Monnard, R.; Seif, J. P.; Tomasi, A.; Jeangros, Q.; Hessler-Wyser, A.; De Wolf, S.; Despeisse, M.; Nicolay, S.; Niesen, B.; Ballif, C. Improved Optics in Monolithic Perovskite/Silicon Tandem Solar Cells with a Nanocrystalline Silicon Recombination Junction. *Advanced Energy Materials* **2018**, *8*. https://doi.org/10.1002/aenm.201701609.

(44) Benzetta, A. E.; Abderrezek, M.; Djeghlal, M. E. Numerical Study of CZTS/CZTSSe Tandem Thin Film Solar Cell Using SCAPS-1D. *Optik* **2021**, *242*. https://doi.org/10.1016/j.ijleo.2021.167320.

(45) Shen, W.; Zhao, Y.; Liu, F. Highlights of Mainstream Solar Cell Efficiencies in 2022. *Frontiers in Energy*. Higher Education Press Limited Company October 2022, pp 9–15. https://doi.org/10.1007/s11708-023-0871-y.

(46) Liu, Z.; Lin, R.; Wei, M.; Yin, M.; Wu, P.; Li, M.; Li, L.; Wang, Y.; Chen, G.; Carnevali, V.; Agosta, L.; Slama, V.; Lempesis, N.; Wang, Z.; Wang, M.; Deng, Y.; Luo, H.; Gao, H.; Rothlisberger, U.; Zakeeruddin, S. M.; Luo, X.; Liu, Y.; Grätzel, M.; Tan, H. All-Perovskite Tandem Solar Cells Achieving >29% Efficiency with Improved (100) Orientation in Wide-Bandgap Perovskites. *Nature Materials* **2025**, *24* (2), 252–259. https://doi.org/10.1038/s41563-024-02073-x.

(47) Geng, C.; Zhang, K.; Wang, C.; Wu, C. H.; Jiang, J.; Long, F.; Han, L.; Han, Q.; Cheng, Y. B.; Peng, Y. Crystallization Modulation and Holistic Passivation Enables Efficient Two-Terminal Perovskite/CuIn(Ga)Se2 Tandem Solar Cells. *Nano-Micro Letters* **2025**, *17*. https://doi.org/10.1007/s40820-024-01514-1.

# Supplementary Information

## Photovoltaic Possibility of $Cu_2SiSe_3$ and $Cu_2SnS_3$ Ternary Chalcogenides: Single Junction to Tandem Architecture

Saptarshi Mandal, Surbhi Ramawat, Sumit Kukreti and Ambesh Dixit*

Advanced Materials and Device (A-MAD) Laboratory, Department of Physics, Indian Institute of Technology Jodhpur, Jodhpur, Rajasthan, 342030 India

*ambesh@iitj.ac.in

### 1. Materials parameters of each layer used for the simulation

Table S1: Material parameters of each layer.

| Parameters | $Cu_2SiSe_3$ [1] | $Cu_2SnS_3$ [2] | CdS [3] | IZO [3] | Al:ZnO [3] |
|---|---|---|---|---|---|
| Thickness (μm) | 1.5 | 1.0 | 0.05 | 0.08 | 0.2 |
| Band-Gap (eV) | 1.44 | 0.91 | 2.42 | 3.37 | 3.37 |
| Electron Affinity (eV) | 4.3 | 4.4 | 4.5 | 4.6 | 4.6 |
| Dielectric Permittivity | 10 | 10 | 9 | 9 | 9 |
| CB Effective DOS ($cm^{-3}$) | $2.2 \times 10^{18}$ | $2.2 \times 10^{18}$ | $1.8 \times 10^{19}$ | $2.2 \times 10^{18}$ | $2.2 \times 10^{18}$ |
| VB Effective DOS ($cm^{-3}$) | $1.8 \times 10^{19}$ | $1.8 \times 10^{19}$ | $2.4 \times 10^{18}$ | $1.8 \times 10^{19}$ | $1.8 \times 10^{19}$ |
| Electron thermal velocity (cm/s) | $10^7$ | $10^7$ | $10^7$ | $10^7$ | $10^7$ |
| Hole thermal velocity (cm/s) | $10^7$ | $10^7$ | $10^7$ | $10^7$ | $10^7$ |
| Electron-Mobility ($cm^2$/Vs) | 100 | 100 | 160 | 150 | 150 |
| Hole-Mobility ($cm^2$/Vs) | 35 | 35 | 50 | 25 | 25 |
| Effective mass of electron | 0.166 | - | 0.25 | 0.275 | 0.275 |
| Effective mass of hole | 0.624 | - | 0.7 | 0.59 | 0.59 |
| Donor-Density ($1/cm^3$) | 0 | 0 | $1 \times 10^{17}$ | $1 \times 10^{17}$ | $1 \times 10^{20}$ |
| Acceptor-Density ($1/cm^3$) | $1 \times 10^{16}$ | $1 \times 10^{17}$ | 0 | 0 | 0 |
| Absorption coefficient | File* | - | SCAPS-1D | SCAPS-1D | SCAPS-1D |
| Coefficient of radiative recombination ($cm^3$/s) | $1.04 \times 10^{-10}$ | $1.04 \times 10^{-10}$ | $1.04 \times 10^{-10}$ | $1.04 \times 10^{-10}$ | - |
| Capture cross-section electron ($cm^2$) | $10^{-15}$ | $10^{-15}$ | $10^{-15}$ | $10^{-15}$ | $10^{-15}$ |
| Capture cross-section hole ($cm^2$) | $10^{-15}$ | $10^{-15}$ | $10^{-13}$ | $10^{-15}$ | $10^{-15}$ |

Table S2: Bulk and interface defect parameters

| Defect Name | $Cu_2SiSe_3$ | $Cu_2SnS_3$ |
|---|---|---|
| Minority carrier lifetime (ns) | $4.9 \times 10^1$ | $4.9 \times 10^1$ |
| Defect type in absorber | Donor | Donor |
| Defect type at Absorber/CdS interface | Neutral | Neutral |
| Recombination speed at Absorber/CdS interface (cm/s) | $10^3$ | $10^3$ |

**2. Contour plots for photovoltaic parameters comparing the thickness and minority carrier lifetime for both $Cu_2SiSe_3$ and $Cu_2SnS_3$ based cell**

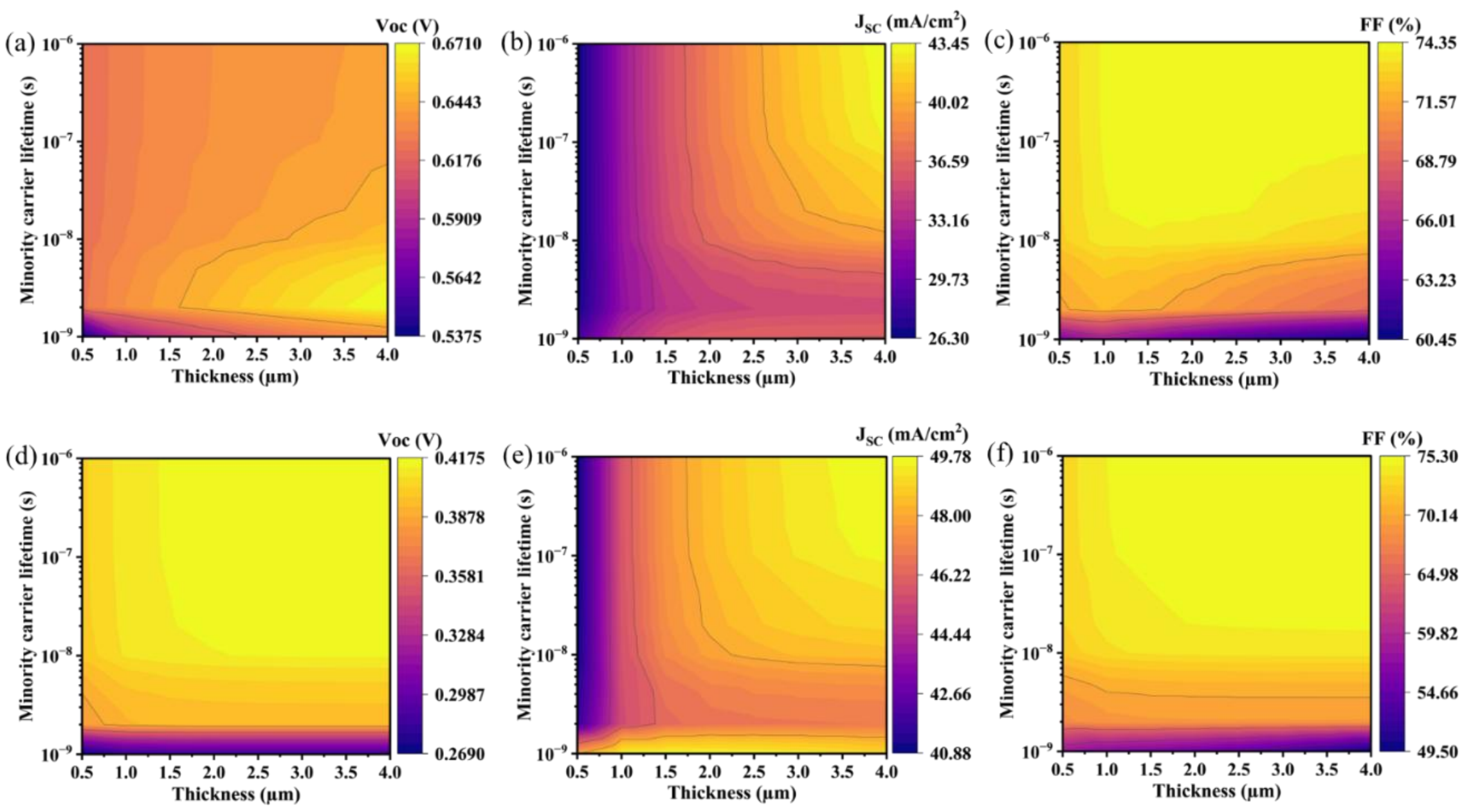

*Figure S1: (a),(b) and (c) showing contour plot of open circuit voltage, short circuit current density and fill factor respectively for different thickness and minority carrier lifetime of the absorber layer for the $Cu_2SiSe_3$ solar cell, and (d),(e) and (f) showing contour plot of open circuit voltage, short circuit current density and fill factor respectively for different thickness and minority carrier lifetime of the absorber layer for the $Cu_2SnS_3$ solar cell*

**3. Variation in the $V_{MPP}$ and $J_{MPP}$ with the carrier concentration for both $Cu_2SiSe_3$ and $Cu_2SnS_3$ based cell**

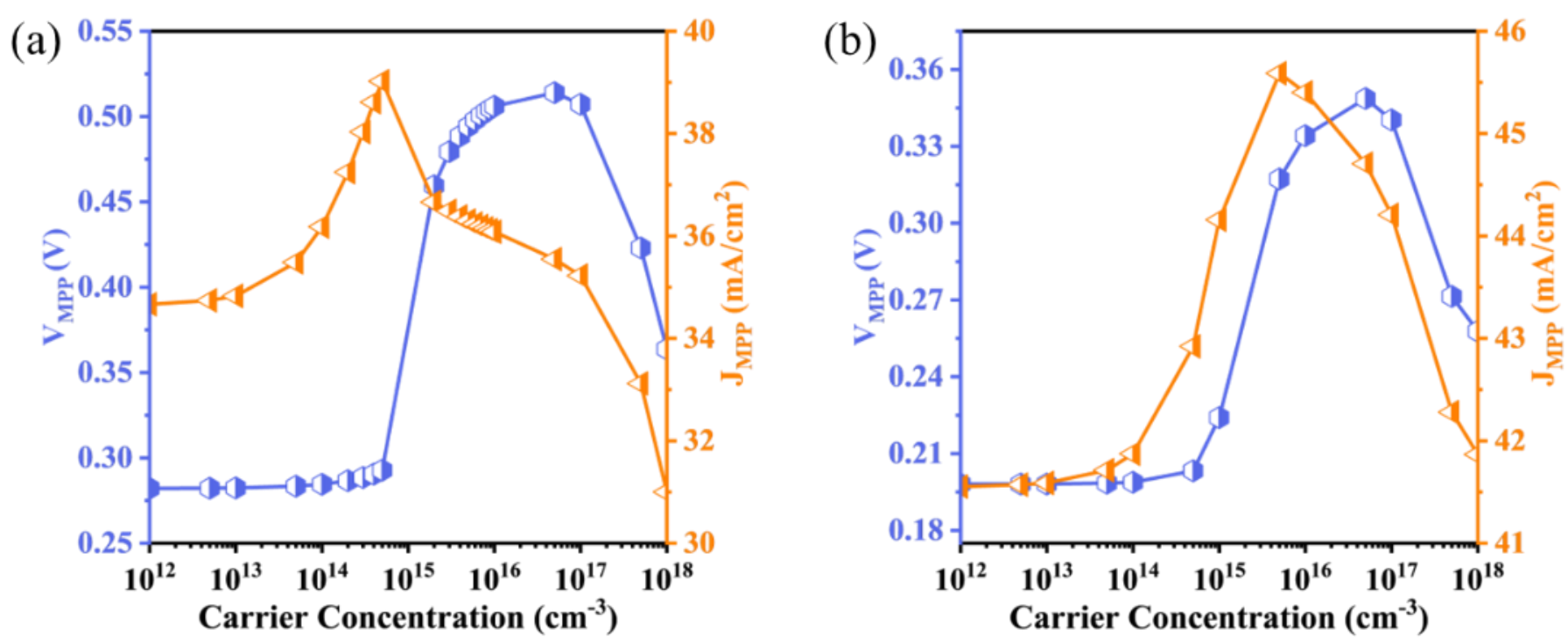

*Figure S2: Variation in $V_{MPP}$ and $J_{MPP}$ with the carrier concentration for (a) $Cu_2SiSe_3$ based solar cell and (b) $Cu_2SnS_3$ based solar cell*

**4. Contour plots for photovoltaic parameters comparing the minority carrier lifetime for both $Cu_2SiSe_3$ and $Cu_2SnS_3$ based cell**

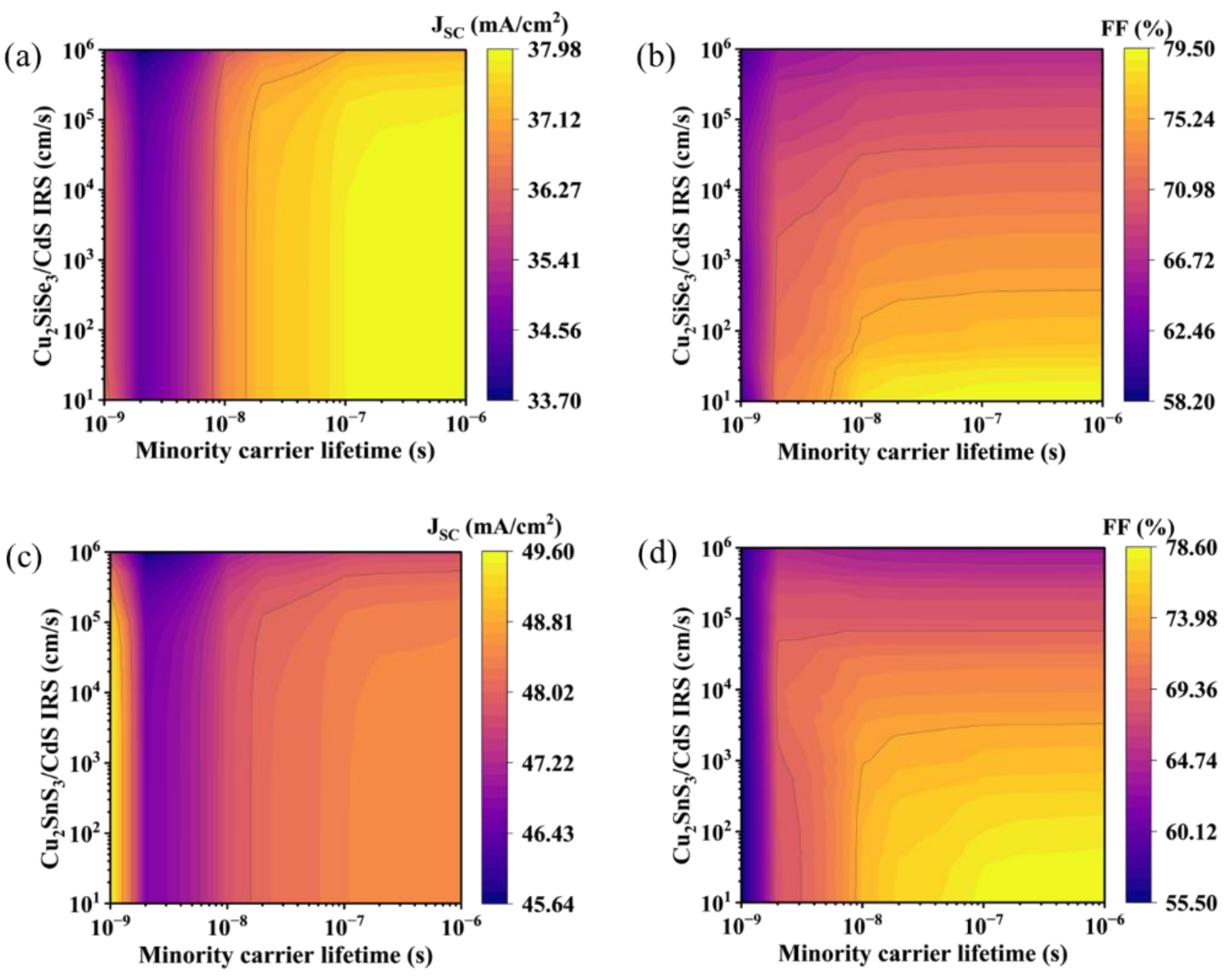

*Figure S3: (a) and (b) showing contour plot of the variation of short circuit current density and fill factor respectively with the minority carrier lifetime and absorber/buffer interface recombination speed (IRS) for the $Cu_2SiSe_3$ solar cell, and (c) and (d) showing contour plot of the variation of short circuit current density and fill factor respectively with the minority carrier lifetime and absorber/buffer interface recombination speed (IRS) for the $Cu_2SnS_3$ solar cell.*

**5. Contour plots for photovoltaic parameters comparing the thickness and carrier concentration of CdS for both $Cu_2SiSe_3$ and $Cu_2SnS_3$ based cell**

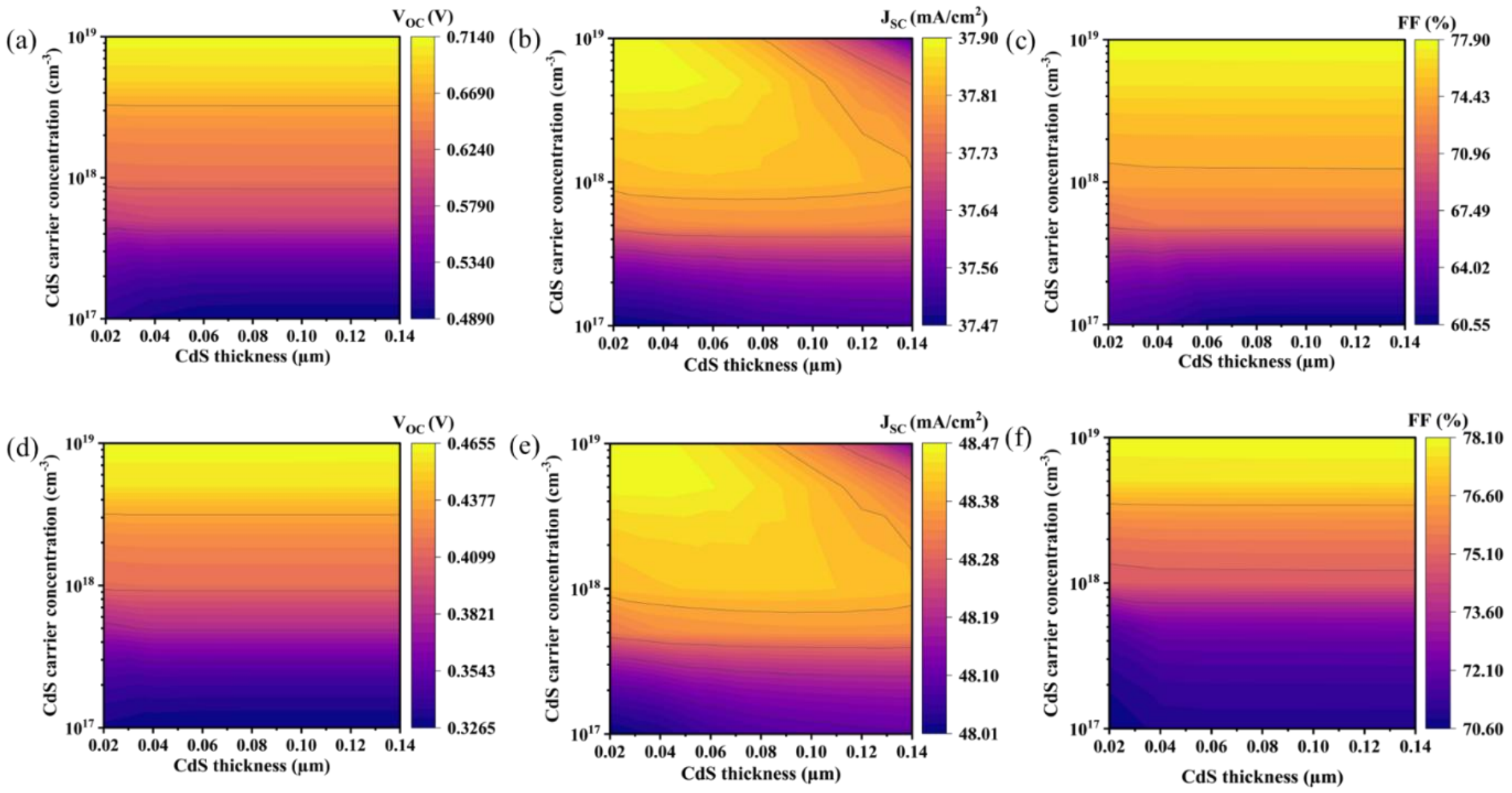

*Figure S4: (a),(b) and (c) showing the contour plots of the variation of open circuit voltage, short circuit current density, and fill factor respectively with CdS carrier concentration and CdS thickness of the $Cu_2SiSe_3$ solar cell, and (d),(e) and (f) showing the contour plots showing the variation of open circuit voltage, short circuit current density, and fill factor respectively with CdS carrier concentration and CdS thickness of the $Cu_2SnS_3$ solar cel.*

**6. Contour plots for photovoltaic parameters comparing the CdS electron affinity and $Cu_2SiSe_3$/CdS interface recombination speed**

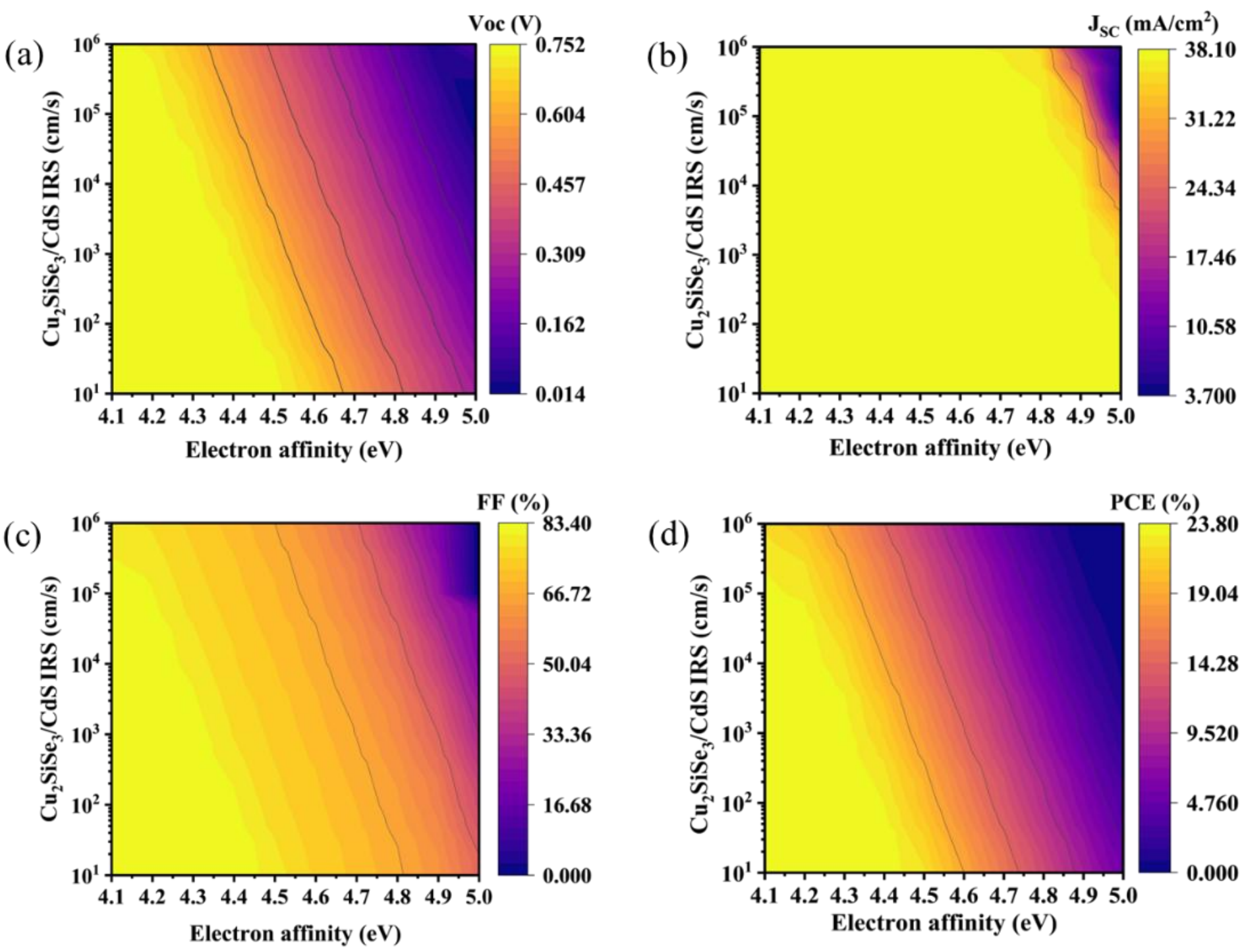

*Figure S5: Contour plot showing the variation of (a) open circuit voltage, (b) short circuit current density (c) fill factor (d) efficiency with absorber/buffer interface recombination speed (IRS) against electron affinity of CdS for $Cu_2SiSe_3$ solar cell.*

## 7. Contour plots for photovoltaic parameters comparing the CdS electron affinity and $Cu_2SnS_3$/CdS interface recombination speed

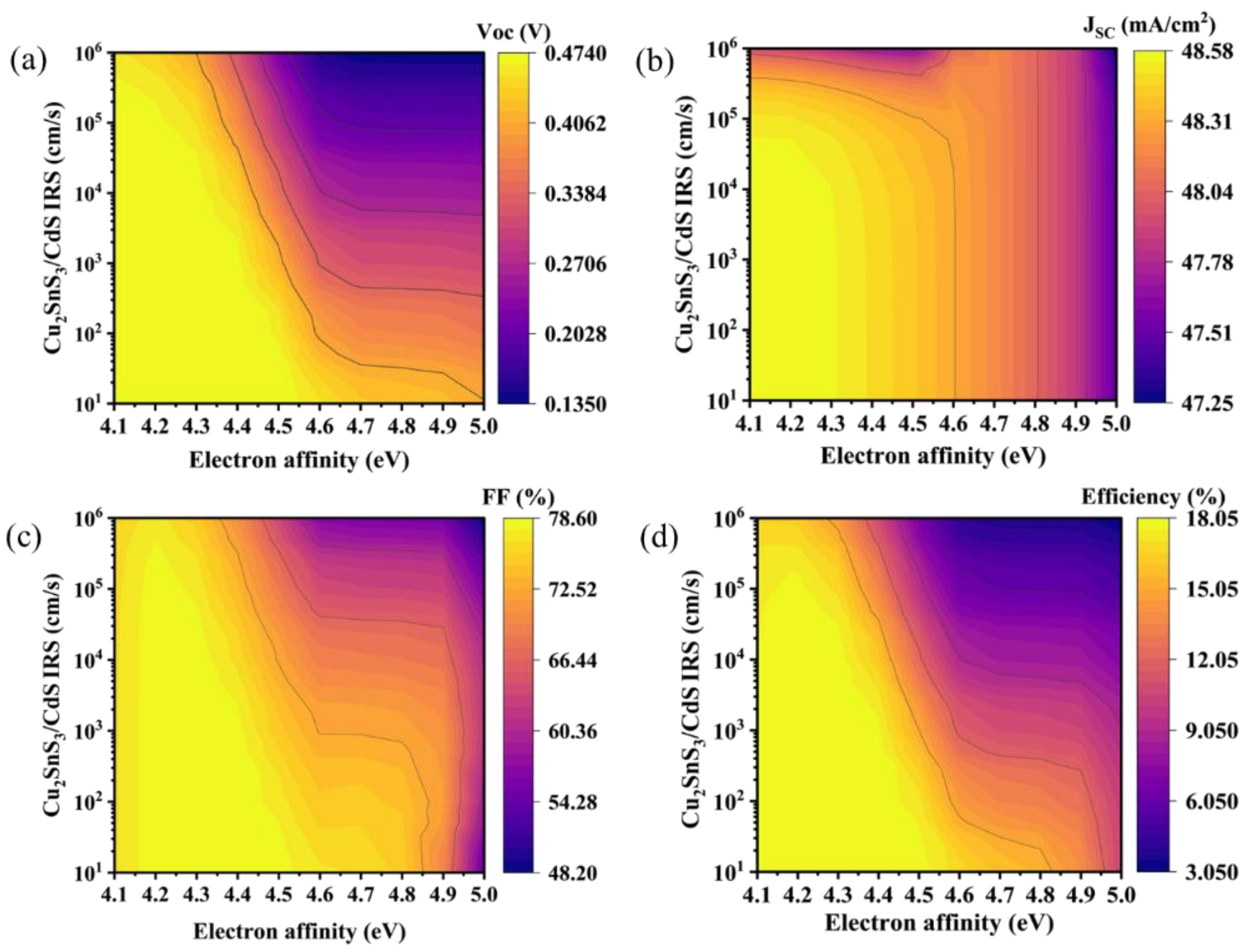

*Figure S6: Contour plot showing the variation of (a) open circuit voltage, (b) short circuit current density (c) fill factor (d) efficiency with absorber/buffer interface recombination speed (IRS) against electron affinity of CdS for $Cu_2SnS_3$ solar cell.*

**8. Contour plot showing the difference in the $J_{SC}$ of top and bottom cell as a function of thickness of absorber of the top and bottom cell.**

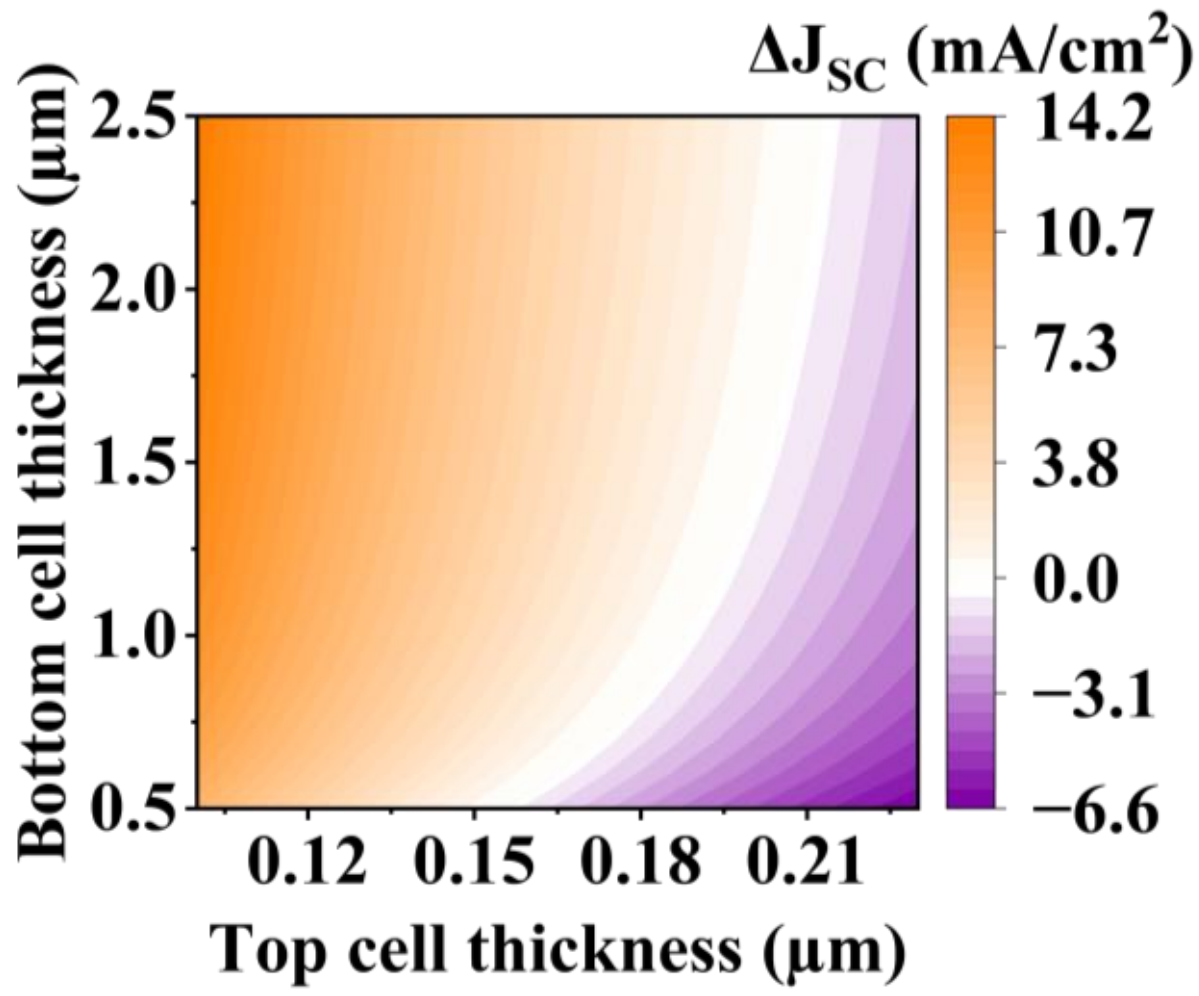

*Figure S7: Contour plot showing variation in $\Delta J_{SC}$ which is the difference in the short circuit current density of top and bottom cell as a function of thickness of the top and bottom cell.*

## References

(1) Kukreti, S.; Ramawat, S.; Dixit, A. Band Anisotropy and Quartic Anharmonicity Cooperate to Drive p - Type Thermoelectricity in the Ternary Diamondlike Semiconductor Cu2SiSe3. *Phys Rev B* **2024**, *110*. https://doi.org/10.1103/PhysRevB.110.245202.

(2) Mathur, A. S.; Upadhyay, S.; Singh, P. P.; Sharma, B.; Arora, P.; Rajput, V. K.; Kumar, P.; Singh, D.; Singh, B. P. Role of Defect Density in Absorber Layer of Ternary Chalcogenide Cu2SnS3 Solar Cell. *Opt Mater (Amst)* **2021**, *119*, 111314. https://doi.org/10.1016/j.optmat.2021.111314.

(3) Kukreti, S.; Gupta, G. K.; Dixit, A. Theoretical DFT Studies of Cu2HgSnS4 Absorber Material and Al:ZnO/ZnO/CdS/Cu2HgSnS4/Back Contact Heterojunction Solar Cell. *Solar Energy* **2021**, *225*, 802–813. https://doi.org/10.1016/j.solener.2021.07.071.